\documentclass[fleqn,usenatbib]{mnras}

\usepackage{newtxtext,newtxmath}
\usepackage[T1]{fontenc}
\DeclareRobustCommand{\VAN}[3]{#2}
\let\VANthebibliography\thebibliography
\def\thebibliography{\DeclareRobustCommand{\VAN}[3]{##3}\VANthebibliography}

\usepackage{epsf}
\usepackage{graphicx}
\usepackage{xspace}    \usepackage{microtype} \usepackage{amsmath}   \usepackage{float}
\usepackage{xcolor} 
\def\ergs {erg~s$^{-1}$}

\DeclareRobustCommand{\msun}{\ensuremath{\mathcal{M}_{\sun}}\xspace}
\DeclareRobustCommand{\mstel}{\ensuremath{\mathcal{M}_*}\xspace}
\DeclareRobustCommand{\mpeak}{\ensuremath{\mathcal{M}_\mathrm{peak}}\xspace}
\DeclareRobustCommand{\mparent}{\ensuremath{\mathcal{M}_\mathrm{parent\;halo}}\xspace}
\DeclareRobustCommand{\msub}{\ensuremath{\mathcal{M}_\mathrm{(sub-)halo}}\xspace}
\DeclareRobustCommand{\mhalo}{\ensuremath{\mathcal{M}_\mathrm{halo}}\xspace}
\DeclareRobustCommand{\LX}{\ensuremath{L_\mathrm{X}}\xspace}
\DeclareRobustCommand{\sar}{\ensuremath{\lambda}\xspace}

\DeclareRobustCommand{\subhalo}{(sub\nobreakdash-)halo\xspace}
\DeclareRobustCommand{\subhaloes}{(sub\nobreakdash-)haloes\xspace}
\DeclareRobustCommand{\umachine}{\textsc{UniverseMachine}\xspace}

\defcitealias{Aird18}{A18}
\newcommand{\PaperII}{\citetalias{Aird18}}

\newcommand{\refone}[1]{{#1}}

\begin{document}
\label{firstpage}
\pagerange{\pageref{firstpage}--\pageref{lastpage}}

\title[The AGN--galaxy--halo connection]{The AGN--galaxy--halo connection: The distribution of AGN host halo masses to $z=2.5$ }
\author[Aird \& Coil]{
James Aird$^{1,2}$\thanks{james.aird@ed.ac.uk} and
Alison L. Coil$^{3}$\\
$^1$Institute for Astronomy, University of Edinburgh, Royal Observatory, Edinburgh EH9 3HJ, UK\\
$^2$School of Physics \& Astronomy, University of Leicester, University Road, Leicester LE1 7RJ, UK\\
$^3$Center for Astrophysics and Space Sciences,
  Department of Physics, University of California, \\
  9500 Gilman Dr. MC 0424,
  La Jolla, CA 92093-0424, USA
}

\date{Submitted to MNRAS on 6th October 2020. Accepted 1st February 2021} 
\pubyear{2021}

\maketitle

\begin{abstract}
It is widely reported, based on clustering measurements of observed active galactic nuclei (AGN) samples, that AGN reside in similar mass host dark matter halos across the bulk of cosmic time, with log~$\mathcal{M}/\mathcal{M}_{\odot}\sim 12.5-13.0$ to $z\sim2.5$. 
We show that this is due in part to the AGN fraction in galaxies rising with increasing stellar mass, combined with  AGN observational selection effects that exacerbate this trend. 
Here, we use AGN specific accretion rate distribution functions determined as a function of stellar mass and redshift for star-forming and quiescent galaxies separately, combined with the latest galaxy-halo connection models, to determine the parent and sub-halo mass distribution function of AGN
\refone{to various observational limits.}
We find that while the median (sub-)halo mass of AGN, $\approx 10^{12}\msun$, is fairly constant with luminosity, specific accretion rate, and redshift, the full halo mass distribution function is broad, spanning several orders of magnitude. 
We show that 
widely used methods
to infer a typical dark matter halo mass based on an observed AGN clustering amplitude can result in biased, systematically high host halo masses.  
While the AGN satellite fraction rises with increasing parent halo mass, we find that the central galaxy is often not an AGN.
Our results elucidate the physical causes for the apparent uniformity of AGN host halos across cosmic time
and underscore the importance of accounting for AGN selection biases when interpreting observational AGN clustering results. 
We further show that AGN clustering is most easily interpreted in terms of the relative bias to galaxy samples, not from absolute bias measurements alone.

\end{abstract}

\begin{keywords}
galaxies: active --
galaxies: evolution --
galaxies: haloes
\end{keywords}

\section{Introduction}
\label{sec:intro}

Accreting supermassive black holes (SMBHs), observed as active galactic nuclei (AGN), are thought to play a crucial role in the galaxy evolution process, influencing the shape of the galaxy stellar mass function \citep[e.g.,][]{Bower12, Puchwein13}, the morphologies of galaxies \citep{Dubois16}, and contributing to the quenching of star formation \citep[e.g.,][]{Croton06, Hopkins06, Zubovas12, Dubois13, Beckmann17}.
AGN feedback in particular is an essential element of modern simulations of galaxy formation and evolution \citep[e.g.,][]{DiMatteo05, Booth09, Richardson16}. 

While the importance of AGN in the lifecycles of galaxies is well established, 
it is not understood  precisely what triggers AGN activity.   The extreme difference in scale between galaxies and supermassive black holes, coupled with the relative rarity of the active accretion phase of SMBHs, has made it difficult to determine the physical mechanisms connecting galaxy and AGN growth. Constraining the triggering and fueling mechanisms of AGNs is crucial to uncovering the relevant physics behind the impact that SMBHs have on their host galaxies.

The spatial distribution or clustering of AGN on large scales is often used to constrain the host dark matter halo properties of AGN and can reveal AGN triggering mechanisms when compared to theoretical models \citep[e.g.,][]{Allevato11, Gatti16}.  Clustering measurements allow AGN to be placed in a cosmological context and reveal  underlying correlations between the large-scale structure of the Universe and AGN fueling.

Most AGN clustering studies, the bulk of which are performed using AGN identified at either optical or X-ray wavelengths,  find that AGN -- whether high-luminosity quasars or more moderate-luminosity AGN -- typically reside in host dark matter halos of mass log $\mhalo/\msun \sim 12.5-13$  from $z\sim0$ to $z\sim4$ (see reviews of X-ray AGN clustering by Cappelluti et al. 2012 and Krumpe et al. 2014 and recent optical quasar clustering results of Timlin et al. 2018 and references therein).
\nocite{Cappelluti12,Krumpe14,Timlin18}
In particular, the typical AGN host halo mass is not found to evolve substantially, if at all, with redshift. 
While small differences in the typical host halo mass have been found when identifying AGN at different wavelengths, luminosities, or obscuration levels \citep[e.g.,][]{Hickox09, Krumpe10,Powell18}, these differences are relatively small when compared with the striking similarity of the typical halo mass over the bulk of cosmic time.  This relative invariance in the halo mass with redshift is difficult to understand, particularly given that the dark matter halo mass of galaxies is a variable function of redshift as well as galaxy properties such as stellar mass and star formation rate \citep[e.g.][and references therein]{Behroozi13,Coil17}. Physically, why would the accretion of mass on the scales of a SMBH at the center of a galaxy be related to a given host halo dark matter mass, and why would the
halo mass of AGN not evolve with redshift in a hierarchical galaxy formation model?

A complicating factor in interpreting AGN clustering results is the strong AGN observational selection effect, 
first shown in \citet{Aird12}, in which SMBHs in more massive galaxies are easier to observe as AGN to a given flux limit of an observational sample.  In essence, more massive galaxies host more massive SMBHs \citep{Cisternas11,Marleau13,Reines15,Simmons17}, therefore to a given flux limit of a survey it is possible to observe AGN at lower Eddington ratios ($L_{\rm AGN}/L_{\rm Edd}$) in more massive galaxies than in lower mass galaxies.  It is simply easier to {\it detect} AGN in more massive galaxies, as they can be accreting at relatively low rates and still be observable, while in lower mass galaxies AGN must be accreting at a high rate relative to the galaxy mass to be observed.   AGN identified in lower mass galaxies will preferentially have high Eddington ratios, as AGN accreting at lower specific accretion rates will be below the flux limit of the observational survey \citep{Aird13}.  When this selection effect is accounted for, it is found that AGN have a wide range of specific accretion rates ($L_{\rm AGN}$/galaxy stellar mass), which roughly translates to a wide Eddington ratio distribution function, and which depends somewhat on the stellar mass of the host galaxy \citep{Georgakakis17,Bongiorno16,Aird18}. 
This selection effect results in AGN being preferentially identified in high mass galaxies \citep[e.g.,][]{Brusa09,Xue10,Mendez13}.

This important selection effect is typically not considered when interpreting AGN clustering results, however. 
Multiple recent AGN clustering studies \citep[][see also \citealt{Leauthaud15}]{Mendez16, Powell18,Georgakakis19} have shown that the clustering of observed AGN samples, whether selected at X-ray, IR, or radio wavelengths, matches the clustering of inactive galaxies of the same stellar mass, star formation rate (SFR) (when calculated), and redshift distributions.  Specifically, this shows that the known galaxy stellar mass to halo mass relation and its scatter---which has been constrained  from galaxy clustering measurements as a function of stellar mass and redshift \citep{Behroozi19}---can be used to infer the dark matter halo mass distribution of AGN host galaxies, once the stellar masses (and ideally SFRs) of AGN host galaxies are determined.  

We show here that the measured broad AGN specific accretion rate distribution function, determined for star-forming and quiescent galaxies as a function of stellar mass and redshift, when combined with the latest empirical models for how galaxies populate dark matter halos \citep[\umachine,][]{Behroozi19},
predicts a roughly constant \refone{median} dark matter halo mass for observed AGN samples across cosmic time.  We predict how the clustering amplitude should vary with AGN luminosity and specific accretion rate, as well as redshift, and compare with current AGN clustering measurements.  
We further 
investigate the differences between the parent and sub-halo mass functions of AGN hosts.  
We also show that the inferred host halo mass derived from observed AGN samples should not be interpreted as reflecting the true AGN host halo mass distribution, which can be lower once the known stellar mass-dependent AGN selection effects are accounted for.
We also predict the satellite fraction as a function of halo mass and show that at high halo masses the central galaxy is often not an AGN.

In Section 2 we describe the model and methodology used in this paper.
We present predictions from this model in Section 3, which we compare with observational results in Section 4. We discuss our findings in Section 5 and present conclusions in Section 6.  Throughout the paper we adopt a flat $\Lambda$CDM cosmology  \citep[$\Omega_m = 0.307, \Omega_\Lambda = 0.693, h = 0.678, \sigma_8 = 0.823, n_s = 0.96$:][]{Planck16}. We evaluate $h$ in all quantities unless explicitly stated.

\section{Modelling methodology}
\label{sec:method}

In this section we describe the methodology we employ to link AGN to dark matter haloes and predict their clustering properties. 
Our method starts from an advanced \emph{galaxy} model \citep[\umachine :][]{Behroozi19}, which uses empirical relations to link the assembly of individual galaxies and their dark matter haloes, constrained by a range of galaxy observables (including stellar mass functions, quenched fractions and galaxy clustering measurements).
We then use recent, robust measurements of the incidence of AGN as a function of galaxy properties \citep{Aird18} to place AGN within these model galaxies and thus---via \umachine---within dark matter haloes. 
This method allows us to predict AGN clustering properties based purely on the known measured relations between AGN and the properties of their host galaxies, combined with the latest galaxy-halo connection models.

\subsection{The \umachine\ galaxy model}

Our starting point is the \umachine\ empirical galaxy model \citep{Behroozi19}. 
Here, we provide a brief summary of the method employed by \umachine\ to allocate galaxies to dark matter haloes, track their evolution over time, predict observables and constrain the model using measurements of galaxy populations. 

For this paper, we adopt \umachine\ data release 1\footnote{\url{http://behroozi.users.hpc.arizona.edu/UniverseMachine/DR1/}} based on the  \refone{\emph{Small MultiDark Planck} (SMDPL)} dark matter simulation \citep{Klypin16,Rodriguez-Puebla16}. 
Dark matter haloes are identified within this large N-body simulation (\refone{400}~$h^{-1}$~Mpc sided co-moving volume with 3840
particles) using the \textsc{RockStar} code \citep{Behroozi13b}, which identifies both individual gravitationally bound haloes and distinct sub-haloes that lie within these parent haloes.
Throughout the remainder of this paper we refer to both individual sub-haloes (containing a satellite galaxy) and the parent haloes (that are associated with a central galaxy) jointly using the specific term ``\subhaloes'' or more generally when using the unqualified term ``halo''.
The merger trees of \subhaloes\ are determined using the \textsc{Consistent Trees} code \citep{Behroozi13d}.

\begin{figure*}
\centering
\includegraphics[width=0.85\textwidth,trim=0 0.5cm 0 0]{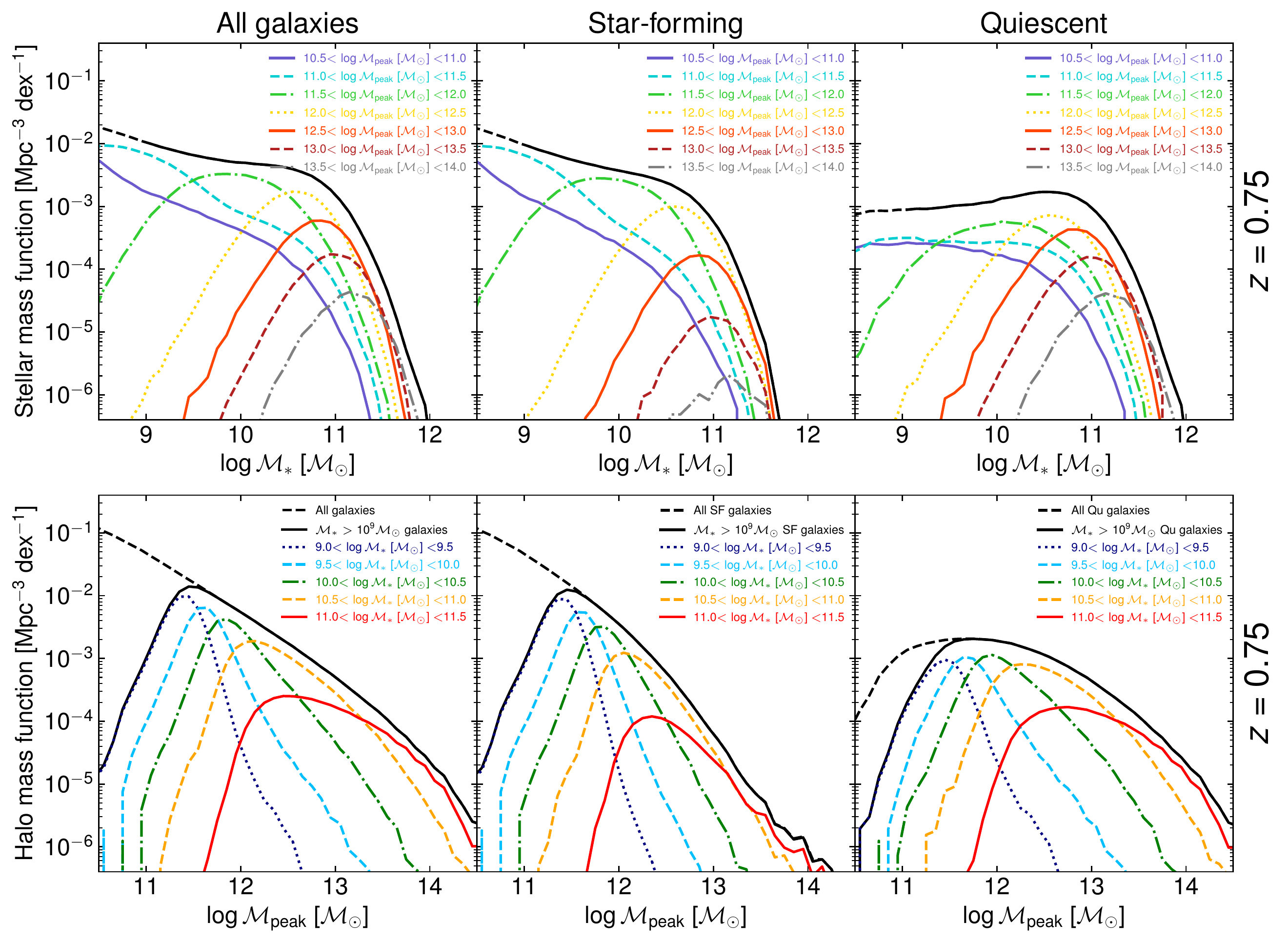}
\caption{Stellar mass functions of galaxies (top panels) and halo mass functions (bottom panels)---defined here as the differential number density of galaxies as a function of peak historical (sub-)halo mass, \mpeak---for all galaxies (left), star-forming galaxies (center) and quiescent galaxies (right), as provided by \umachine\ \citep{Behroozi19}, shown here at $z=0.75$.
In the top panels, we show the contribution to the total stellar mass function from galaxies in haloes with different \mpeak\ by the colored lines.
In the bottom panels, we show the contribution to the total halo mass function from galaxies with different stellar masses by the colored lines.
The thick black line in the bottom panel indicates the halo mass function of our ``parent galaxy sample" used in this work, which is restricted to galaxies with $\mstel >10^{9}\msun$. 
This cut introduces a turnover in the halo mass functions of all galaxies and star-forming galaxies at $\mhalo\sim 10^{11}\msun$ (cf. the total halo mass function including lower \mstel\ galaxies shown by the dashed black line).
We note that higher \mstel\ galaxies have a broad range of \mpeak\ and thus make a significant contribution to the total halo mass function across a wide range of \mstel. When considering only star-forming galaxies (centre panels) there is a closer mapping between \mpeak\ and \mstel, although the intrinsic scatter in the SMHM relation still produces a somewhat broad distribution of \mstel\ for galaxies of given \mpeak. 
}
\label{fig:smfs_and_hmfs}
\end{figure*}

Galaxies are then associated with these individual \subhaloes---and the stellar mass assembly of the galaxy is traced---using empirical relations to assign SFRs.
The distribution of SFRs is defined for haloes of a given $v_{\mpeak}$ (the  maximum circular velocity of the halo, $v_\mathrm{max}$, at its peak historical halo mass, \mpeak) and redshift.
SFRs for a given $v_{\mpeak}$ and $z$ are modelled as a bimodal distribution corresponding to star-forming galaxies and some fraction of ``quenched'' galaxies, where the quenched fraction also depends on both $v_{\mpeak}$ and $z$. 
Individual haloes (at a given $v_{\mpeak}$) are ranked based on the relative growth of the halo  (traced by the change in $v_\mathrm{max}$) over the previous dynamic time, and SFRs are assigned based on this ranking such that haloes that are currently growing faster are assigned higher SFRs at the current redshift,\footnote{
In satellite galaxies whose sub-haloes may be undergoing strong tidal stripping, the change in $v_\mathrm{max}$ is instead taken over the period since the sub-halo reached its peak historical mass, \mpeak\ (if this exceeds the dynamical time).  
Changes in $v_\mathrm{max}$ tend to occur suddenly for such galaxies, followed by longer periods (exceeding the dynamical time) where $v_\mathrm{max}$ remains relatively stable. Thus, adopting this longer time period prevents such satellite galaxies being artificially assigned high SFRs and ensures that they remain quenched.
See section~3.2 of \citet{Behroozi19} for full details.
}
allowing for scatter in this relation.
The total stellar mass growth of individual galaxies over cosmic time is  determined by integrating the allocated SFRs over cosmic time and through the mergers of individual \subhaloes.

Finally, the model is used to predict a range of galaxy observables, including (but not limited to) stellar mass functions,
cosmic star formation rates,
specific SFR distributions,
quenched fractions,
and autocorrelation functions for star-forming and quiescent galaxies across cosmic time. 
Observational errors and systematic effects are added before comparing the predictions to real data and determining a $\chi^2$ for a given set of input parameters. 
The input parameters---a total of 44 free parameters controlling the assembly of galaxies within haloes---are then explored using a Markov Chain Monte Carlo (MCMC) algorithm to determine the best-fit empirical model.

The output of the \umachine\ model links halo masses to the current stellar mass (\mstel) and SFR of individual galaxies, providing robust predictions of the stellar mass--halo mass (SMHM) relation of different galaxy populations and---most crucially---the underlying \emph{scatter} in these relations. 
In Figure~\ref{fig:smfs_and_hmfs} we demonstrate the relationship between stellar mass and 
\subhalo\ mass, as determined by \umachine.
Here, and throughout the remainder of this paper, we adopt the ``observed'' stellar mass of galaxies that is calculated by \umachine, including the effects of observational errors that can have a systematic impact on the recovered properties. Adopting the observed stellar mass ensures we can directly link the \umachine\ output to other observables (i.e.~AGN fractions as a function of \mstel).\footnote{Adopting the true stellar mass estimates from \umachine\ instead of the observed values \refone{marginally reduces the breadth of the distributions of stellar masses at fixed halo mass shown in Figure~\ref{fig:smfs_and_hmfs}} but does not have a significant impact on our results or overall conclusions.}
We use the observationally-determined division of star-forming and quiescent galaxies as adopted by \citet{Aird18} when measuring AGN fractions (see Section~\ref{sec:agnfrac} below). Specifically, we define quiescent galaxies as having
\begin{equation}
\log \mathrm{SFR}\; [\msun\;\mathrm{yr}^{-1}] < -8.9 + 0.76 \log \frac{\mstel}{\msun}   + 2.95\log(1+z). 
\end{equation}
This cut cleanly divides the \umachine\ model galaxy populations for the range of stellar masses ($\mstel>10^{9}\msun$) considered in this paper.

The top row of Figure~\ref{fig:smfs_and_hmfs} shows the overall stellar mass
functions (SMFs) of all galaxies and separately for the star-forming and quiescent galaxy populations (shown here at $z=0.75$ only, for demonstration purposes). 
In each panel we show the contributions to the SMFs from galaxies in haloes with different masses. 
The bottom row shows the halo mass functions (HMFs) for each galaxy population and the contributions from galaxies with different stellar masses.
We adopt the peak historical \subhalo\ mass, \mpeak, which 
traces the overall growth of the halo in which a given galaxy grows 
and is thus the key quantity to consider for understanding galaxy evolution \citep{Reddick13}.
Figure~\ref{fig:smfs_and_hmfs} shows how \subhaloes\ of a given \mpeak\ host galaxies with a broad range of stellar masses; correspondingly, galaxies of a given \mstel\ are found in haloes with a broad range of \mpeak. 
When considering star-forming galaxies only there is a closer mapping between \mstel\ and \mpeak, although the distribution of \mstel\ for a given \mpeak\ remains fairly broad. 
For star-forming galaxies with $\mstel=10^{10-10.5}\msun$, the width of the halo mass function that encloses 90\% of galaxies is $\sim$0.6~dex, but increases toward higher stellar masses reaching $\sim$1.1~dex for galaxies with  $\mstel=10^{11-11.5}\msun$. 
For quiescent galaxies, the distribution of \mpeak\ (at a given \mstel) is broader, spanning $\sim$0.9~dex for galaxies with $\mstel=10^{10-10.5}\msun$ and increasing to $\sim$1.5~dex for galaxies with $\mstel=10^{11-11.5}\msun$. 
Simpler schemes---for example using existing SMHM relations for all galaxies to map stellar masses to halo masses directly---would not capture this diversity of \mpeak\ across the galaxy population
and the differing host haloes of star-forming and quiescent galaxies that is described by the \umachine\ model.

\subsection{Populating galaxies with AGN}
\label{sec:agnfrac}

The next stage of our method is to populate the model galaxies from \umachine\ with AGN. 
To achieve this, we take a purely empirical approach, adopting measurements of the incidence of X-ray selected AGN for samples of galaxies by \citet[hereafter \PaperII]{Aird18}, either taking all galaxies (of given \mstel\ and $z$) or considering star-forming and quiescent galaxies as distinct populations. 
\refone{We use these observations to populate galaxies in the \umachine\ snapshot closest to the central redshift of the bins used by \PaperII.}\footnote{
\refone{Specifically, we take the snapshots at scale factor $a=0.766904$, 0.571997, 0.445435, 0.364435 and 0.308748 that correspond to redshifts $z=0.3$, 0.75, 1.25, 1.75 and 2.25. We do not explore higher redshifts where the AGN fraction is poorly determined by the \PaperII\ observations.}}

\PaperII\ took large, near-infrared--selected samples of galaxies from the CANDELS and UltraVISTA surveys \citep{Grogin11,Skelton14,McCracken12,Muzzin13} and extracted X-ray data at the positions of every galaxy from deep \textit{Chandra} imaging \citep{Alexander03,Xue11,Nandra15,Civano16}.
A Bayesian methodology was used to combine the X-ray data from samples of galaxies at particular ranges of \mstel\ and $z$ and recover robust estimates of the underlying distribution of specific accretion rates, $\lambda$, defined as 
\begin{equation}
    \lambda = \frac{ k_\mathrm{bol} \LX }
                   { 1.3\times 10^{38}\;\mathrm{erg\;s^{-1}} \times 0.002 \frac{\mstel}{\msun} }
    \label{eq:lambda}    
\end{equation}
where \LX\ is the rest-frame 2--10~keV luminosity (based on \mbox{2--7}~keV observed X-ray fluxes assuming an unabsorbed spectrum with photon index $\Gamma=1.9$) and $k_\mathrm{bol}$ is a bolometric correction (a constant $k_\mathrm{bol}=25.0$ was assumed). 
The denominator of Equation~\ref{eq:lambda} ensures that $\lambda\approx$ the Eddington ratio, under the assumption that $0.002\mstel\approx\mathcal{M}_\mathrm{BH}$, the mass of the central black hole in a galaxy of stellar mass \mstel.
Constant scaling factors are chosen such that $\lambda\propto\frac{\LX}{\mstel}$, which means $\lambda$ can be described as the specific accretion rate, the amount of black hole growth relative to the mass of the galaxy, and is thus independent of underlying uncertainties or scatter in the \mstel--$\mathcal{M}_\mathrm{BH}$ relation. 
As discussed in Section~\ref{sec:intro}, adopting measurements in terms of specific accretion rate accounts for the \mstel-dependent selection bias in AGN samples, whereby an AGN in a more massive galaxy produces a higher observable luminosity and is thus easier to detect than AGN in less massive galaxies.
In addition, the Bayesian methodology accurately accounts for the varying sensitivity of the X-ray observations---both within and between different fields---and probes down to $\sim$an order of magnitude below the nominal sensitivity limits of the \textit{Chandra} imaging \citep[see also][]{Aird17}.

In this paper, we take the \PaperII\ measurements of $p(\lambda\;|\; \mstel,z)$, which describes the probability that a galaxy with a given stellar mass and redshift hosts an AGN as a function of $\lambda$, and integrate them down to fixed limits in $\lambda$ to provide estimates of the \emph{fraction} of AGN in a given galaxy sample \citep[see also][]{Aird19}. 
Such estimates provide a robust definition of the AGN fraction in terms of the number of galaxies that are growing their black holes above a certain threshold in $\lambda$, i.e.~relative to the mass of the galaxy. 
In addition, we reverse Equation~\ref{eq:lambda} to recover estimates of the AGN fraction to  various limits in \LX. These estimates account for the varying sensitivity of the X-ray observations but do not allow for the broad \mstel-dependent selection biases. As such, they provide a robust estimate of the \emph{observed} AGN fraction that allows for known instrumental effects but does not account for intrinsic selection biases in the sample.

\begin{figure*}\includegraphics[width=0.85\textwidth, trim=0 0.5cm 0 0 ]{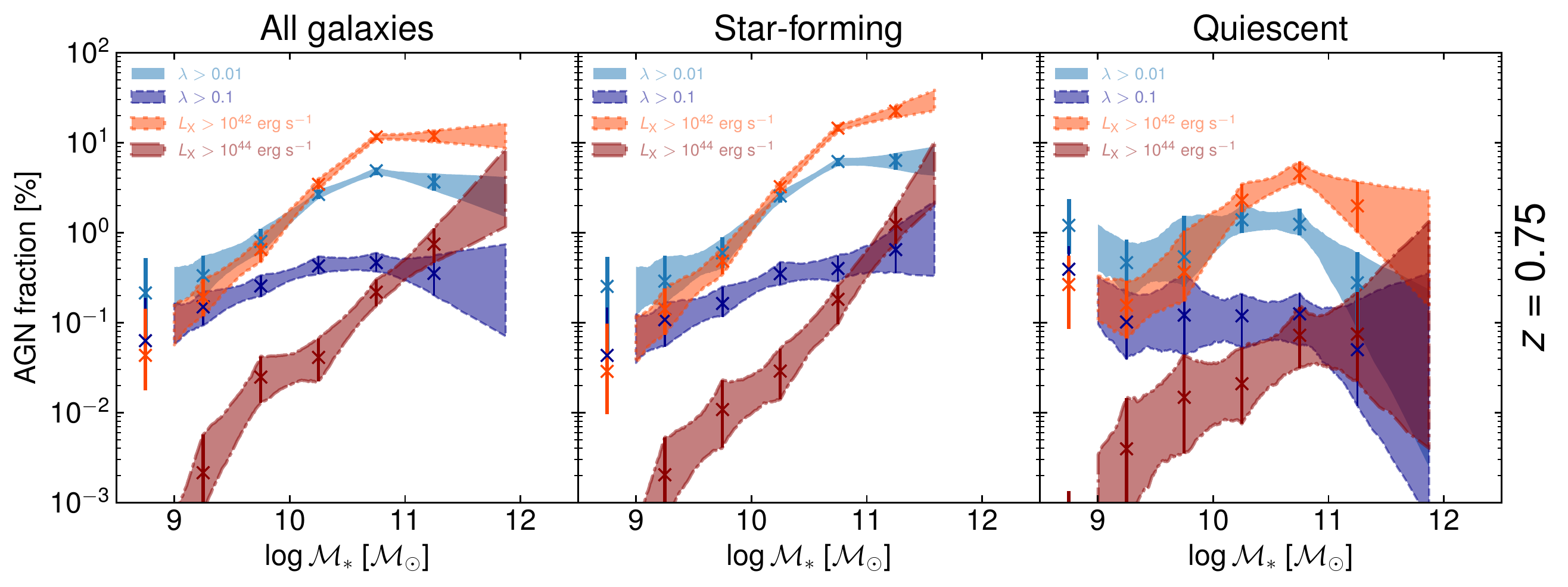}
\caption{
AGN fraction (shown here at $z=0.75$) as a function of \mstel\ in all galaxies (left), star-forming galaxies (center) and quiescent galaxies (right) for four ``AGN samples" defined to different selection limits in terms of X-ray luminosity (\LX) or specific accretion rate ($\lambda\propto\LX/\mstel$).
Crosses mark the robust measurements of the AGN fraction within observed galaxy samples from \citet{Aird18} along with $1\sigma$-equivalent uncertainties. The colored regions indicate our interpolation (or extrapolation) of the AGN fraction to the \mstel\ of each galaxy in the \umachine\ sample, with the width of the region indicating the $1\sigma$-equivalent uncertainties that are propoagated from the original measurements.
We note the mild \mstel\ dependence of the AGN fraction to $\lambda$ limits in all and star-forming galaxies that becomes more exaggerated when adopting \LX\ limits.
In quiescent galaxies the AGN fraction to $\lambda$ limits is almost constant with \mstel; again a \mstel\ dependence is introduced when adopting \LX\ limits  and neglecting the observational bias against finding growing black holes in lower \mstel\ galaxies.
}
\label{fig:agnfractions}
\end{figure*}

Figure~\ref{fig:agnfractions} shows the AGN fractions (shown here at $z=0.75$) that we calculate based on the \PaperII\ measurements for all galaxies (of given \mstel\ and $z$; left panel) and considering star-forming and quiescent galaxies as separate populations with a distinct AGN fraction as a function of \mstel\ and $z$ within each population (middle and right panels). 
To determine the AGN fraction within our \umachine\ model galaxies, we linearly interpolate between the centres of the \mstel\ bins from \PaperII\ to the values of each individual galaxy (extrapolating for
$\mstel>10^{11.5}\msun$).
We use the underlying posterior distributions of the $p(\lambda\;|\;\mstel,z)$ measurements and propagate these uncertainties into our estimates of the AGN fraction for \umachine\ galaxies. 
Rather than creating a mock catalogue of AGN, we instead weight each galaxy by the AGN fraction, retaining the full fidelity of the underlying \umachine\ galaxy sample and allowing us to efficiently propogate uncertainties in the AGN fraction.
Shaded regions in Figure~\ref{fig:agnfractions} (and subsequent figures) encompass the 68\% central confidence interval based on the propagated posterior distributions.

As shown in Figure~\ref{fig:agnfractions}, the AGN fraction to fixed $\lambda$ limits in all galaxies, or when considering star-forming galaxies only, is found to rise with increasing \mstel, i.e.~there is an intrinsic \mstel\ dependence such the AGN activity is more common in higher \mstel\ galaxies, even after accounting for observational selection biases. 
The trend is slightly weaker when considering a higher $\lambda>0.1$ limit for the AGN fraction (dark blue region), indicating that high-$\lambda$ activity does not depend so strongly on the galaxy mass. 
The \mstel-dependent trends are exaggerated when considering samples to limits in \LX\ due to the observational selection bias discussed above. 
For quiescent galaxies, the AGN fraction (to $\lambda$ limits) is approximately constant with \mstel, with tentative evidence for a \emph{decrease} at $\mstel>10^{11}\msun$. 
Adopting \LX\ limited samples introduces an artificial \mstel\ dependence due to the selection bias. 
\refone{The same pattern is observed in all our redshift slices, although the stellar-mass dependence for $\lambda$-limited AGN samples in star-forming galaxies is somewhat milder for our lowest redshift slice ($z=0.3$, see \PaperII).}

Throughout this paper, we do not consider AGN in galaxies with $\mstel<10^{9}\msun$ or their host haloes. 
Measurements of the AGN fraction in such low-mass galaxies are highly uncertain, especially at $z\gtrsim0.7$ \citep[although see e.g.~][for recent measurements at lower $z$]{Mezcua18,Birchall20}. 
Given the higher space densities of such low-mass galaxies, the large uncertainties in the AGN fraction in this regime begin to dominate recovered quantities such as average halo masses (e.g. Section~\ref{sec:interpret_bias}). 
We thus limit our analysis to a parent sample of galaxies with $\mstel>10^9\msun$. 
This mass limit introduces a low-mass turnover in the corresponding HMFs (see solid black lines in the bottom panels of Figure~\ref{fig:smfs_and_hmfs}) 
that is propagated to our AGN HMFs (see Section~\ref{sec:agn_hmfs_and_wp}). 
We note that observed AGN samples (e.g.~X-ray selected, \LX-limited samples) are dominated by AGN in galaxies with $\mstel\gtrsim10^{10}\msun$ \citep{Xue10,Aird13}.
In addition, only the relatively rare, high-$\lambda$ AGN in lower mass galaxies ($\mstel\lesssim10^9\msun$) would satisfy the luminosity limit of $\LX>10^{42}$~\ergs\ typically adopted for X-ray AGN samples. 
Thus our model will provide a good description of observed AGN samples and enables us to make robust statements regarding the true properties of AGN within the $\mstel>10^9\msun$ galaxy population.

We note that being based on 2--7~keV observed energies, the \PaperII\ measurements will include both unobscured and moderately obscured AGN but will not include the contribution from heavily obscured, Compton-thick sources. 
Under the assumption that the intrinsic fraction of Compton-thick fraction of AGN does not strongly depend on $\lambda$ \citep[e.g.][]{Ricci17}, our measurements can be used to infer the properties (e.g.~halo mass functions) of \emph{all} AGN by applying an appropriate scaling factor ($\sim1.25-2.0$) to $p(\lambda\;|\;\mstel,z)$ and the derived AGN fractions, SMFs and HMFs.\footnote{Assuming no intrinsic dependence of the Compton-thick fraction on host galaxy properties, our predicted clustering measurements and average halo masses should not change.} 
However, given remaining uncertainties on the overall Compton-thick fraction \citep[ranging from $\sim20-50$\%:][]{Ueda14,Ricci15,Buchner15,Aird15,Ananna19} and any intrinsic dependence on host galaxy properties \citep[e.g.][]{Koss11,Kocevski15} we do not apply such a correction here and restrict our analysis and conclusions to the radiatively efficient (i.e.~X-ray luminous), Compton-thin AGN population.

Our method shares some similarities with recent work by \citet{Georgakakis19}, who used measurements of AGN specific accretion rate distributions from \PaperII\ and complementary work by \citet{Georgakakis17} to populate mock galaxy catalogues with AGN \citep[see also][]{Comparat19}.
While the conclusions of \citet{Georgakakis19} are consistent with our model, we focus on a number of different aspects of the resulting halo mass functions and interpretation of the clustering properties of AGN. 
We use a more sophisticated model to link galaxies and their dark matter haloes, via \umachine, that directly models the distinct halo masses of star-forming and quiescent galaxy populations, whereas \citet{Georgakakis19} adopt a single SMHM relation (with scatter) and separate by galaxy type in post-processing,
thus neglecting systematic differences in the host haloes of these populations. 
In addition, we show explicitly how observational selection biases impact the results and we provide a detailed analysis of how to interpret measurements of AGN clustering and bias.

\section{Model predictions of AGN halo mass functions and clustering properties}
\label{sec:agn_hmfs_and_wp}

We now use the empirically motivated model detailed in the previous section to predict the HMFs of AGN samples selected to different \LX\ observational limits and more complete AGN samples selected to a range of $\lambda$ limits. 
We also use our model to predict the clustering properties of AGN samples and compare these to the underlying galaxy population. 

\refone{
In this section, we present results primarily at $z=0.75$, where our measurements of the AGN fraction are best constrained. 
However, the \LX\ and $\lambda$ dependence, explored in Sections~\ref{sec:lxlims} and \ref{sec:lambdalims},  is similar in the other redshift slices that we investigate. 
In Section~\ref{sec:lxlambda_comparison} we compare the clustering strength of our AGN samples at different redshifts.
We defer discussion of the redshift evolution of the HMF itself to Section~\ref{sec:parent_vs_sub} below.}

\begin{figure*}    \centering
    \includegraphics[width=0.63\textwidth]{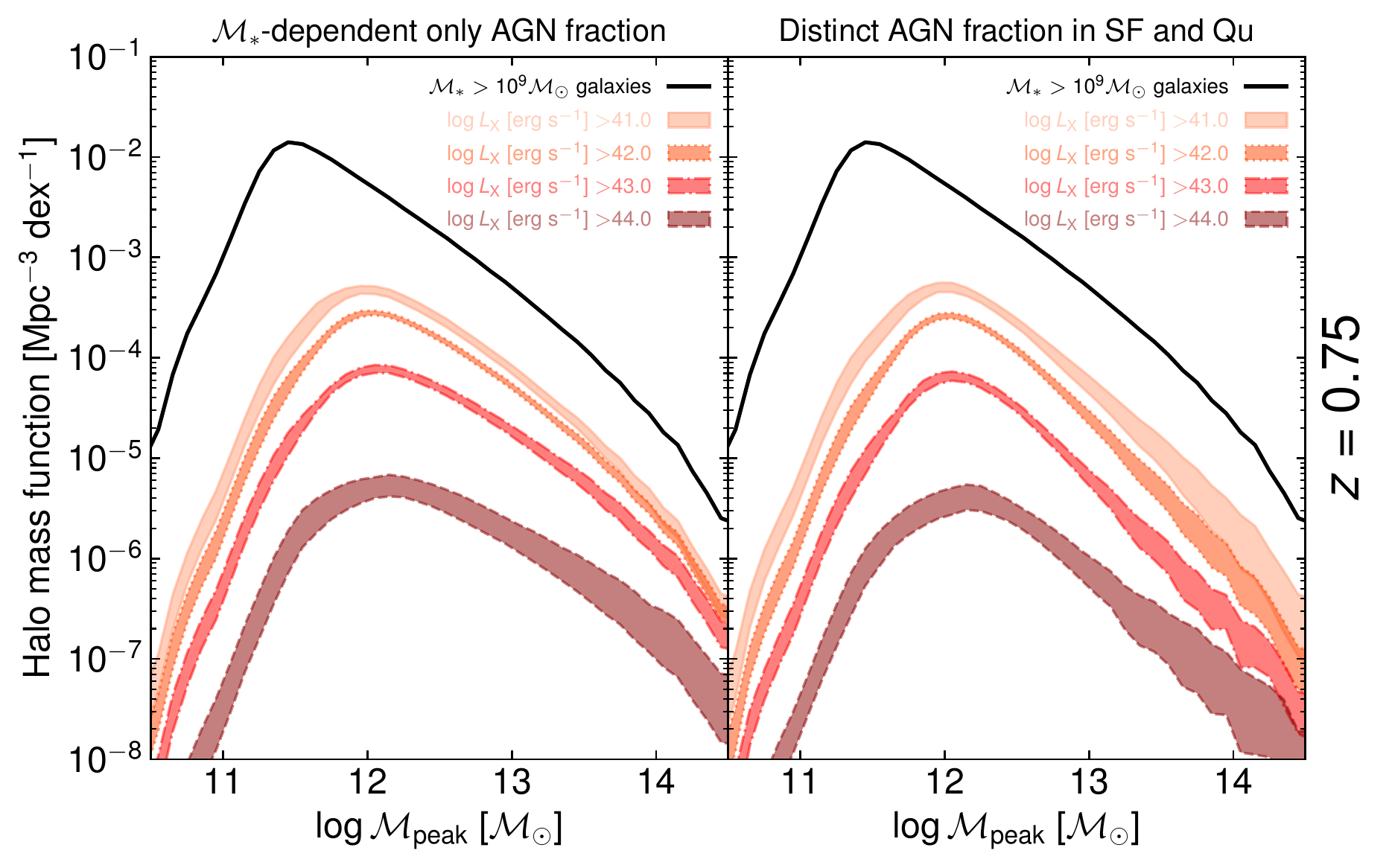}
    \caption{
    Halo mass functions (differential number density as a function of \mpeak) for a range of \LX-limited AGN samples, as indicated by different coloured regions, using our two different methods to populate galaxies with AGN. \textbf{Left panel:} using measurements of the AGN fraction in all galaxies as a function of \mstel\ only. \textbf{Right panel:} placing AGN in galaxies depending on their classification (star-forming or quiescent) and using the distinct measurements of the AGN fraction (as a function of \mstel) in each population. 
    Adopting an \mstel-dependent only AGN fraction tends to place a higher number of AGN in high-\mstel\ quiescent galaxies, which tend to be found in more massive (higher \mpeak) haloes. 
    Adopting the distinct measurements of the AGN fraction in star-forming and quiescent galaxies preferentially assigns AGN to massive star-forming galaxies and thus suppresses the high-\mpeak\ tail of the AGN halo mass function. 
    For both of our methods, the AGN halo mass function reaches a maximum at slightly higher \mpeak\ than the underlying $\mstel>10^{9}\msun$ parent galaxy population (solid black line), reflecting the increased fraction of AGN (to a given \LX\ limit) in higher \mstel\ galaxies.
    }
    \label{fig:hmfs_lxlims}
\end{figure*}

\subsection{\LX-limited AGN samples}
\label{sec:lxlims}

Figure~\ref{fig:hmfs_lxlims} shows the predicted HMFs of AGN samples selected to various \LX\ limits, comparing our two different approaches to populate galaxies with AGN.
In the left panel, we use measurements of the AGN fraction in all galaxies, adopting the observed \mstel\ dependence only (see left panel of Figure~\ref{fig:agnfractions}). 
In the right panel of Figure~\ref{fig:hmfs_lxlims} we  adopt the distinct measurements of the AGN fraction (as a function of \mstel) in star-forming and quiescent galaxies (see middle and right panels of Figure~\ref{fig:agnfractions}).
The recovered HMFs with both modelling approaches show that \LX-limited samples of AGN are predicted to be found in haloes with a broad range of \mpeak, regardless of the adopted luminosity limit.
For the \mstel-dependent only model, 
the width of the HMF that contains 90\% of AGN is $\sim$1.6~dex for $\LX>10^{42}$~\ergs\ AGN, increasing to $\sim$1.9~dex for $\LX>10^{44}$~\ergs\ AGN.
Adopting a distinct AGN fraction in star-forming and quiescent galaxies suppresses the number of AGN found in higher mass haloes ($\mpeak\gtrsim10^{12}\msun$), reducing the  the width containing 90\% of AGN to $\sim$1.4--1.7~dex. 
These results
differ from the 
\mstel-dependent only model (shown in the left panel) as AGN are preferentially assigned to star-forming galaxies, that tend to lie in lower \mpeak\ haloes, and AGN are suppressed in quiescent galaxies, that tend to lie in more massive haloes. 
For both models, the AGN HMFs peak at $\mpeak\approx10^{12}\msun$, approximately 0.5~dex higher than the underlying $\mstel>10^9\msun$ parent galaxy population (black solid lines), reflecting the strong preference of \LX-limited AGN samples to be found in higher \mstel\ galaxies, due to both intrinsic effects and observational biases.

\begin{figure*}
    \centering
    \includegraphics[width=0.63\textwidth]{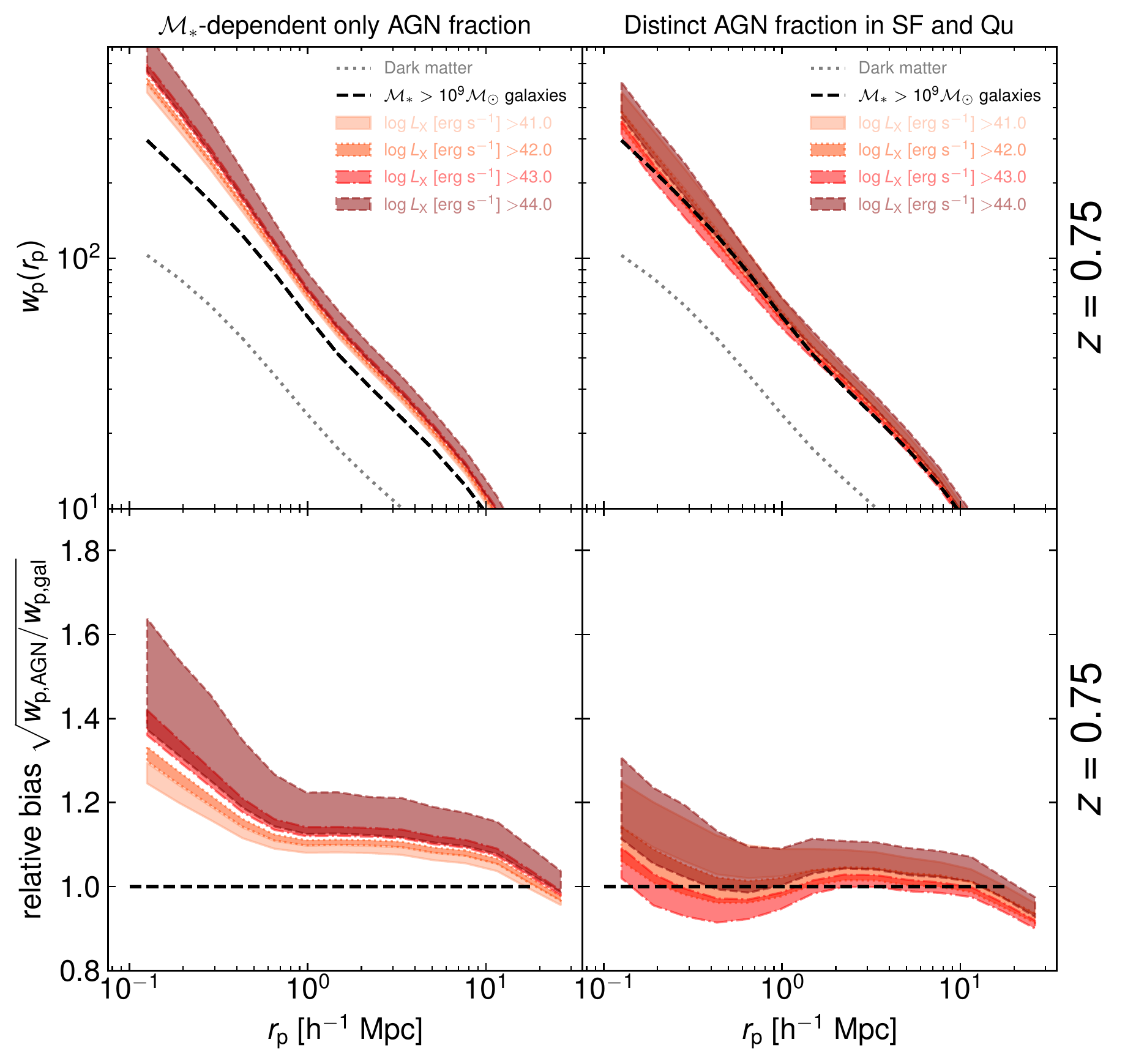}
    \caption{
    \textbf{Top:} Measurements of the projected clustering amplitude as a function of projected distance, $w_p(r_p)$, for a range of \LX-limited AGN samples (as indicated by the coloured regions), using our two different methods to populate galaxies with AGN (left versus right panels, as indicated).
    The black dashed lines show $w_p(r_p)$ of the parent $\mstel>10^{9}\msun$ galaxy sample, while the grey dotted lines show $w_p(r_p)$ of dark matter, including non-linear effects \citep[based on][]{Mead16}. 
    \textbf{Bottom:} Relative bias of the AGN samples as a function of $r_p$, compared to the parent  $\mstel>10^9\msun$ galaxy sample. 
    }
    \label{fig:wp_Lxlims}
\end{figure*}

\begin{figure}
    \centering
    \includegraphics[width=0.9\columnwidth]{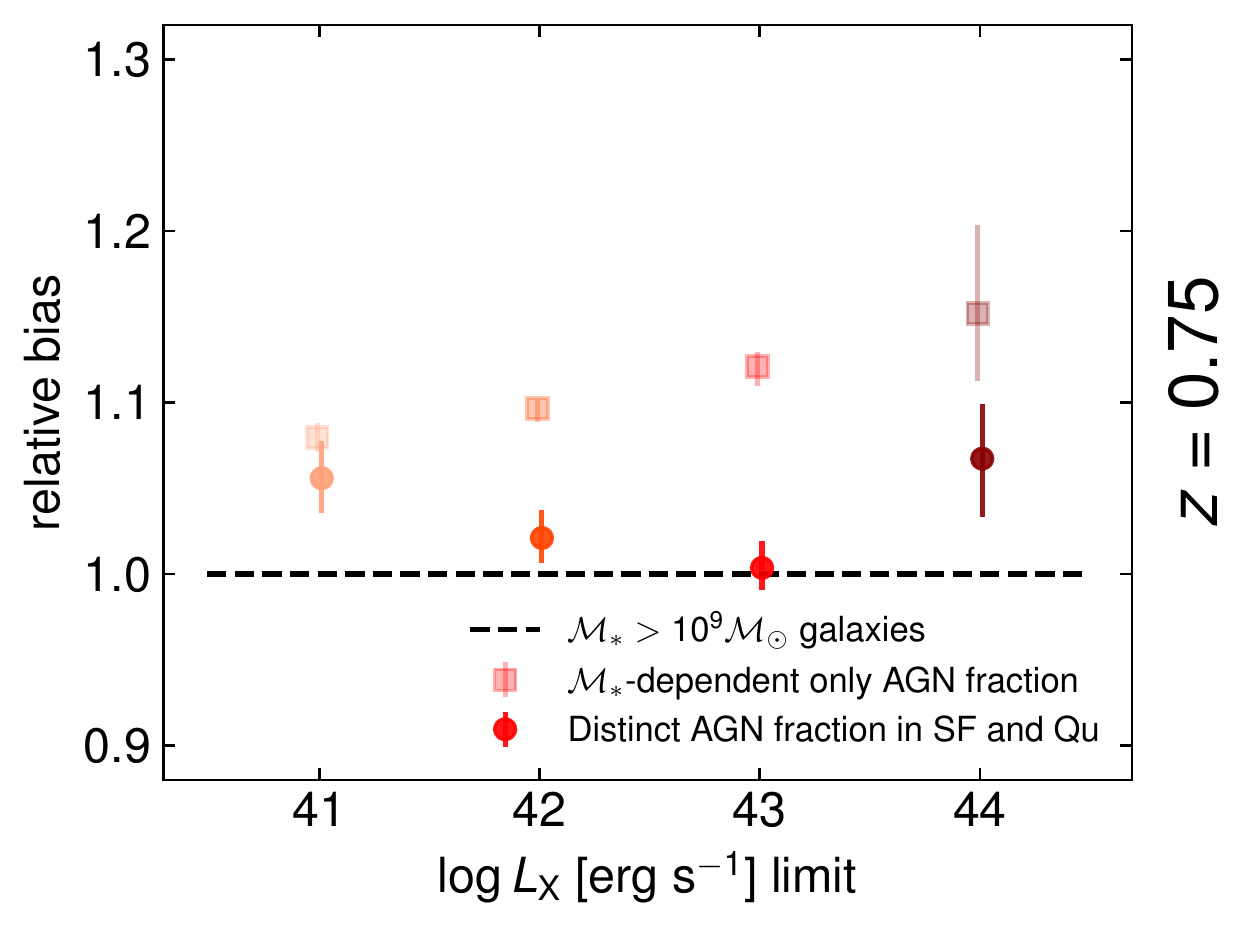}
    \caption{Relative bias of AGN compared to the parent  $\mstel>10^9\msun$ galaxy sample as a function of the \LX\ limit, averaged over scales of $1 h^{-1}$~Mpc~$< r_p < 10 h^{-1}$~Mpc, comparing our two different methods to populate galaxies with AGN (squares and circles, as indicated). 
    The relative bias shows a mild dependence on \LX\ for the \mstel-dependent only AGN fraction; the relative bias is generally lower and the \LX\ dependence is weaker when adopting the distinct AGN fractions in star-forming and quiescent galaxies. 
    }
    \label{fig:bias_vs_lxlim}
\end{figure}

We also 
predict the \emph{observable} clustering properties of \LX-limited AGN samples 
\refone{for a given simulation snapshot.}
Following the majority of observational work (see Section 4 below), we determine projected two-dimensional auto-correlation functions,
\begin{equation}
    w_p(r_p) = 2 \int_0^{\pi_\mathrm{max}} \xi(r_p,\pi) d\pi
\end{equation}
where $\xi(r_p,\pi)\equiv\xi(r)$, the two-point correlation function as a function of co-moving distance $r$, which we separate into the projected on-sky distance, $r_p$ \refone{(calculated from the $x$ and $y$ positions in co-moving Mpc in a given \umachine\ snapshot)}, and line-of-sight distance, $\pi$ \refone{based on the $z$ position in \umachine.}
Integrating $\xi(r_p,\pi)$ in the $\pi$ direction accounts for redshift-space distortions to the true positions of galaxies and their AGN in real observational datasets. 
We adopt $\pi_\mathrm{max}=60 h^{-1}$~Mpc.

To predict realistic $w_p(r_p)$ measurements using our model, we first add redshift-space distortions to \refone{the positions of} our \umachine\ model galaxies based on \refone{the peculiar velocities of their haloes in the $z$ direction \citep[using \textsc{HaloTools} v0.7,][]{Hearin17}.}
We then measure $w_p(r_p)$ \refone{from the co-moving co-ordinates} using the \textsc{CorrFunc} code \citep{Sinha20}, which accounts for the weighting of each \umachine\ galaxy by the appropriate AGN fraction.
The efficiency of the \textsc{CorrFunc} code enables us to repeat our calculations of $w_p(r_p)$ for each of the posterior draws in our AGN fraction and thus fully propagate these uncertainties.

The top panels of Figure~\ref{fig:wp_Lxlims} show our model predictions for $w_p(r_p)$ for our various \LX-limited AGN samples. 
We also show the corresponding $w_p(r_p)$ of the underlying $\mstel>10^9\msun$ galaxy population (black dashed line), which is also calculated from \umachine\ using \textsc{CorrFunc}, and for dark matter particles (grey dotted line) which is calculated from the matter power spectrum, including non-linear effects, given by \citet{Mead16}. 
The bottom panels of Figure~\ref{fig:wp_Lxlims} show the relative bias of the AGN samples relative to the $\mstel>10^9\msun$ galaxy population as a function of scale.  It is assumed that
\begin{equation}
    w_{p,\mathrm{AGN}}(r_p) = b_\mathrm{rel}^2(r_p)\; w_{p,\mathrm{gal}}(r_p)
\end{equation}
where $w_{p,\mathrm{AGN}}$ and $w_{p,\mathrm{gal}}$ are the projected two-point correlation functions of the AGN sample and parent galaxy sample, respectively, and $b_\mathrm{rel}$ is the relative bias. 
With the \mstel-dependent only AGN fraction, we find the amplitude of $w_p(r_p)$ for the \LX-limited samples is significantly higher than the parent galaxy sample, reflecting the typically higher \mpeak\ of the host haloes (as seen in the HMFs shown in Figure~\ref{fig:hmfs_lxlims}). 
We predict a lower amplitude for the relative bias when adopting a distinct AGN fraction in star-forming and quiescent galaxies, reflecting the reduction in the AGN incidence at higher \mpeak\ seen in Figure~\ref{fig:hmfs_lxlims}. 
Figure~\ref{fig:bias_vs_lxlim} shows the average of the relative bias over scales of 
\mbox{$1 h^{-1}$~Mpc $< r_p < 10 h^{-1}$~Mpc} as a function of the \LX\ limit of the AGN samples, comparing our two model approaches. 
The \mstel-dependent only model predicts a mild rise in the relative bias as the \LX\ limit of the sample increases.
Our more realistic model that adopts a distinct AGN fraction in star-forming and quiescent galaxies (as observed) predicts an approximately {\it constant} relative bias, with tentative evidence of an increase for the highest luminosities ($\LX>10^{44}$~\ergs). 
In both cases we predict a mild---but significant---enhancement of the measured bias of \LX-limited AGN samples by $\sim$3--15\% compared to $\mstel>10^{9}\msun$ galaxies, reflecting the typically higher halo masses of X-ray selected AGN samples and their enhanced clustering properties.

\begin{figure*}
    \centering
    \includegraphics[width=0.63\textwidth]{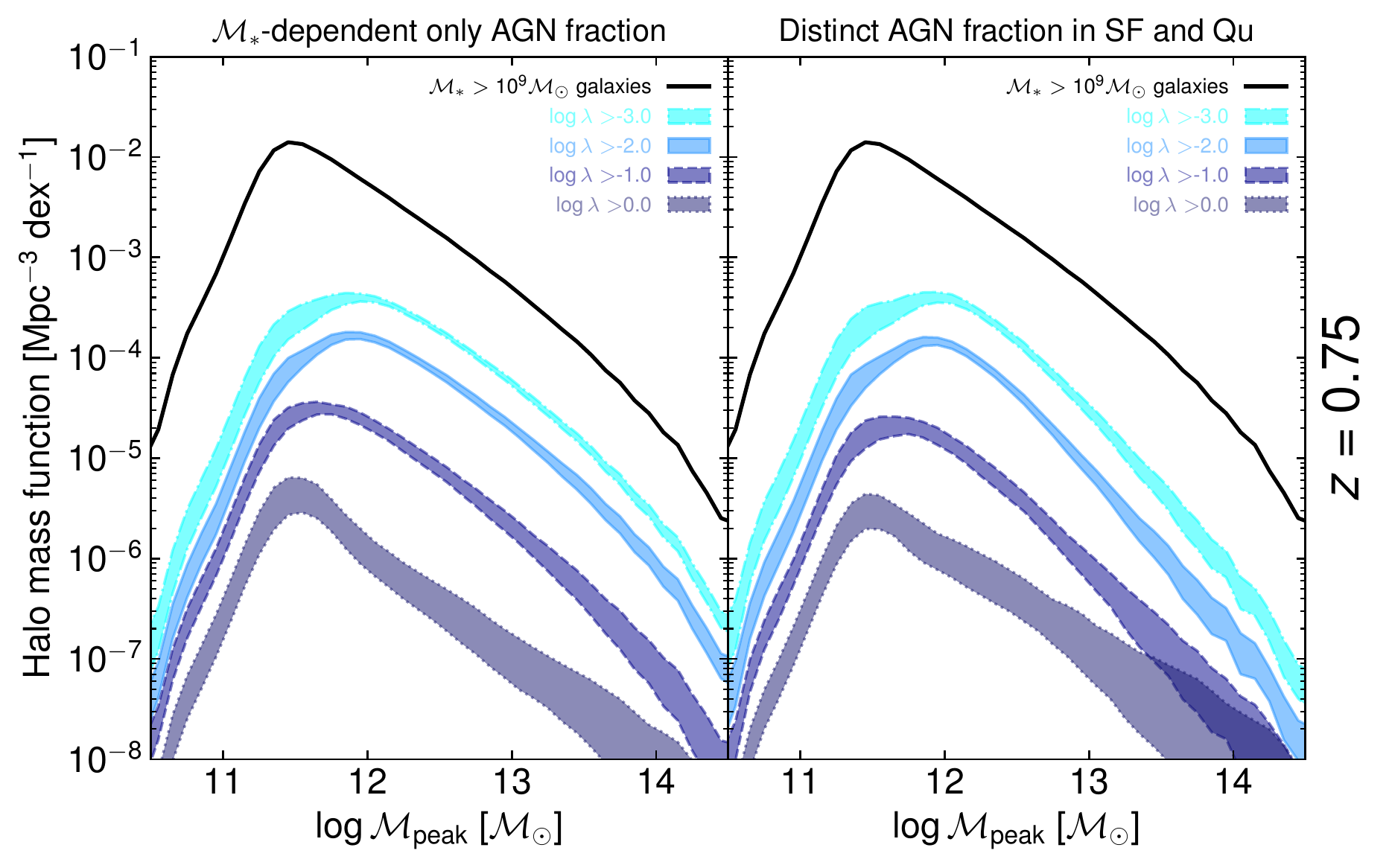}
    \caption{
       Halo mass functions (differential number density of AGN as a function of \mpeak) for a range of $\lambda$-limited AGN samples, as indicated by different coloured regions, using our two different methods to populate galaxies with AGN. 
       For higher $\lambda$ limits the shape of the halo mass function approaches the shape of the underlying $\mstel>10^{9}\msun$ galaxy sample (black line)---albeit with a much lower normalisation---reflecting the weaker \mstel\ dependence of the AGN fraction for such samples. 
    }
\label{fig:hmfs_lambdalims}
\end{figure*}

\begin{figure}
    \centering
    \includegraphics[width=0.9\columnwidth]{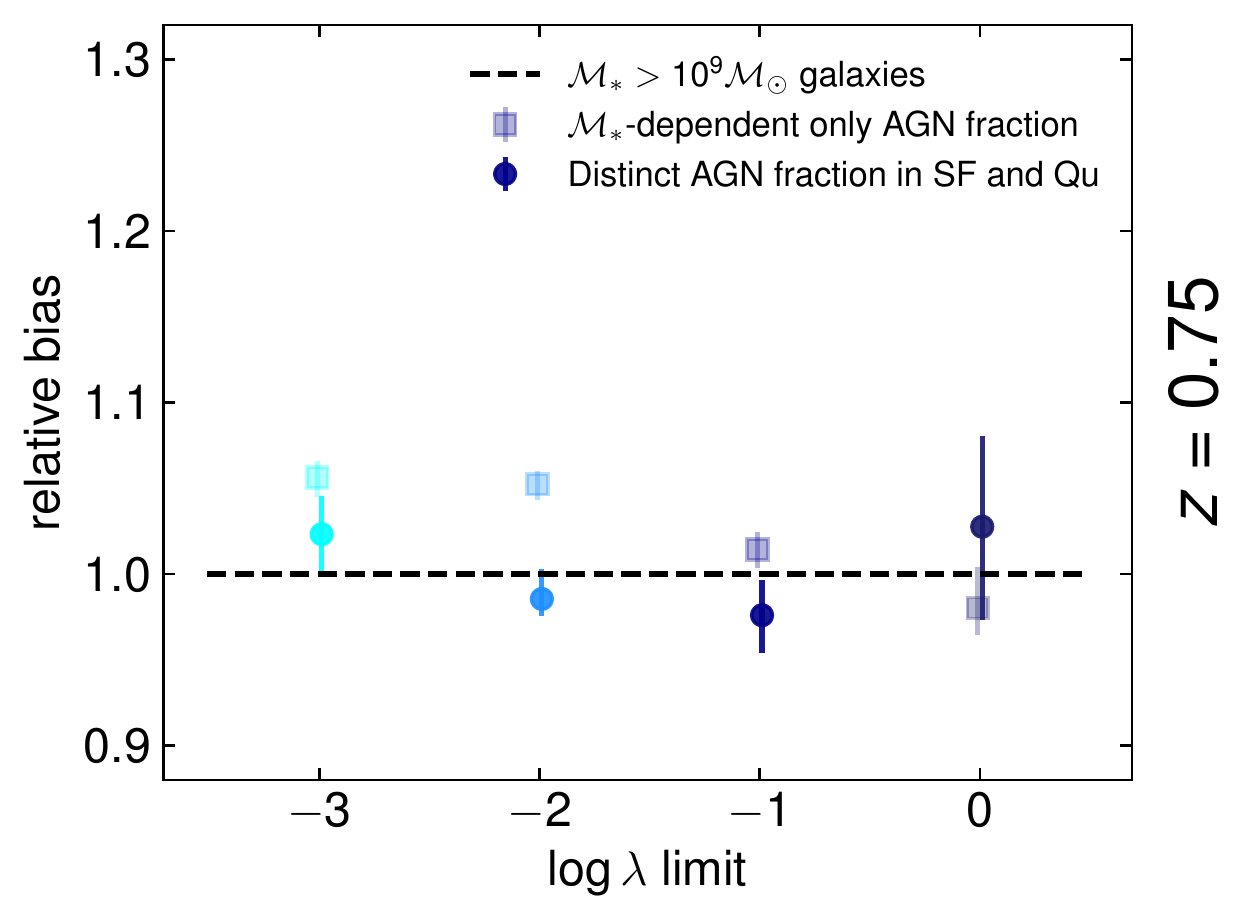}
    \caption{Relative bias of AGN compared to the parent  $\mstel>10^9\msun$ galaxy sample as a function of the $\lambda$ limit, averaged over scales of $1 h^{-1}$~Mpc $< r_p < 10 h^{-1}$~Mpc, comparing our two different methods to populate galaxies with AGN. 
    A mild \emph{decrease} in the relative bias with increasing $\lambda$ is found for the \mstel-dependent AGN fraction (squares). When assigning AGN independently in star-forming and quiescent galaxies we find the relative bias is consistent with no dependence on $\lambda$ and at $z=0.75$ is consistent with the parent galaxy population.
    }

    \label{fig:bias_vs_lambdalim}
\end{figure}

\subsection{$\lambda$-limited AGN samples}
\label{sec:lambdalims}

Selecting AGN to a fixed limit in luminosity introduces an observational
bias toward higher stellar mass hosts, and thus toward higher halo masses. 
Here, we instead present HMFs and the clustering properties of AGN samples selected to fixed limits in specific accretion rate, $\lambda$. 
Adopting $\lambda$ limits allows us to identify galaxies as AGN based on how rapidly they are growing their black holes, countering the observational selection bias toward higher \mstel\ hosts and providing estimates of the \emph{underlying}
distribution of AGN halo masses,\footnote{As our analysis is restricted to galaxies with $\mstel>10^9\msun$, a more accurate statement is that we present the underlying distribution of AGN halo masses \emph{within this parent galaxy sample}.} 
in contrast to \emph{observed} samples of AGN explored in Section~\ref{sec:lxlims} above.

In Figure~\ref{fig:hmfs_lambdalims} we present HMFs for $\lambda$-limited AGN samples, using our two different methods to populate galaxies with AGN. 
As for the \LX-limited samples, the HMFs of $\lambda$-limited AGN samples are broad, with 90\% of the AGN spanning $\gtrsim$1.4~dex in \mpeak. 
Adopting our more accurate model, where we use a distinct AGN fraction in star-forming and quiescent galaxy populations, reduces the predicted number density of AGN at higher halo masses ($\mpeak\gtrsim10^{12}\msun$) as fewer AGN are placed in high-\mstel\ quiescent galaxies, which generally inhabit higher \mpeak\ haloes.
For both models, the peak of the HMF moves toward lower \mpeak\ as the $\lambda$ limit is increased. 
This trend is due to the differences in the measured \mstel-dependence of the fraction of AGN for different $\lambda$ limits. 
The incidence of moderate accretion rate black holes (e.g. $\lambda>0.01$) is enhanced in higher stellar mass ($\mstel\gtrsim10^{10.5}\msun$) star-forming galaxies, whereas the incidence of more rapidly accreting black holes does not show this mass dependence and thus the AGN HMFs approach the same shape as the underlying parent galaxy population (albeit with a much lower normalisation). 
Such behaviour is also seen in the simpler \mstel-dependent only model.

We also derive $w_p(r_p)$ for our \sar-limited samples as described in Section~\ref{sec:lxlims} above. 
In Figure~\ref{fig:bias_vs_lambdalim} we present the relative bias (compared to the parent galaxy population, averaged over \mbox{$1 h^{-1}$~Mpc $< r_p < 10 h^{-1}$~Mpc}) as a function of the \sar\ limit for the two model approaches. 
For our most accurate model (adopting the distinct AGN fractions in star-forming and quiescent galaxies, as observed), we predict a relative bias that is consistent with 1 (at $z=0.75$), regardless of \sar\ limit, {\it consistent with there being no observable difference in the clustering properties of AGN and the underlying galaxy population.}

\subsection{Comparison between \LX- and $\lambda$-limited AGN samples and the underlying galaxy population}
\label{sec:lxlambda_comparison}

\begin{figure}
\centering
\includegraphics[width=0.8\columnwidth]{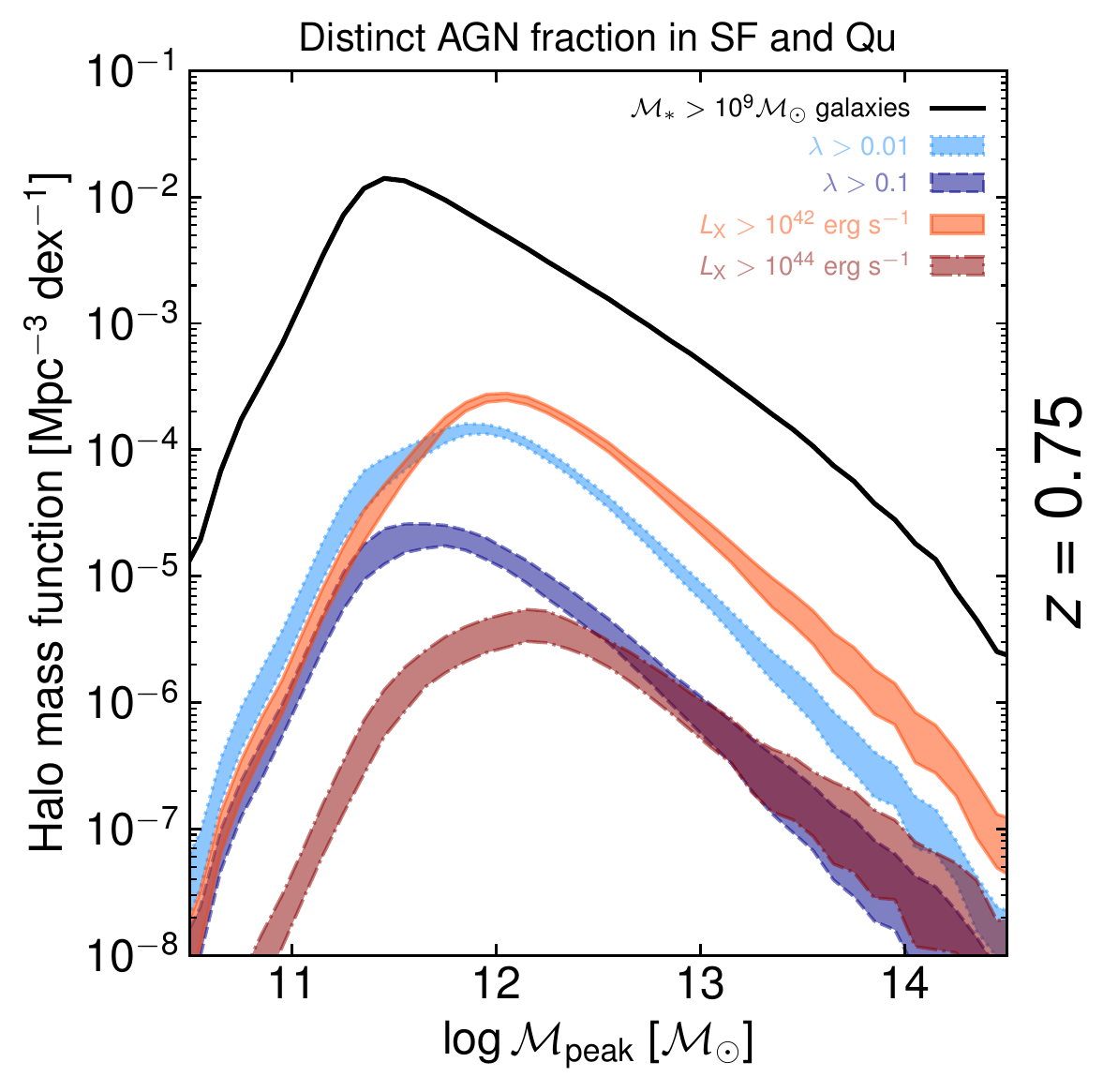}
\caption{
Halo mass functions for four ``AGN samples" to different \LX\ and $\lambda$ limits, using our preferred method to populate galaxies with AGN (using the distinct measurements of AGN fraction in star-forming and quiescent galaxies). 
The halo mass functions for $\lambda$-limited samples---that account for \mstel-dependent AGN selection biases---peak at lower \mpeak\ and more closely track the halo mass function of the underlying galaxy population. 
}
\label{fig:hmfs_foursamples}
\end{figure}

In Figure~\ref{fig:hmfs_foursamples} we directly compare the HMFs for  \LX-limited and \sar-limited AGN samples as predicted by our more sophisticated and accurate model (adopting a distinct AGN fraction in star-forming and quiescent galaxies).
Differences in the overall normalisation reflect differences in the fraction of haloes that contain AGN, depending on the definition, with higher \LX\ and higher \sar\ samples thus corresponding to lower overall normalisations.
However, there are also significant differences in \emph{shapes} of the HMFs, seen most clearly in the comparison between the high-luminosity AGN sample ($\LX>10^{44}$~\ergs; dark red) and the 
complete
sample of highly accreting black holes ($\sar>0.1$; dark blue) which are found in significantly greater numbers in lower mass haloes ($\mpeak<10^{12}\msun$). 

\begin{figure}
    \centering
    \includegraphics[width=0.8\columnwidth]{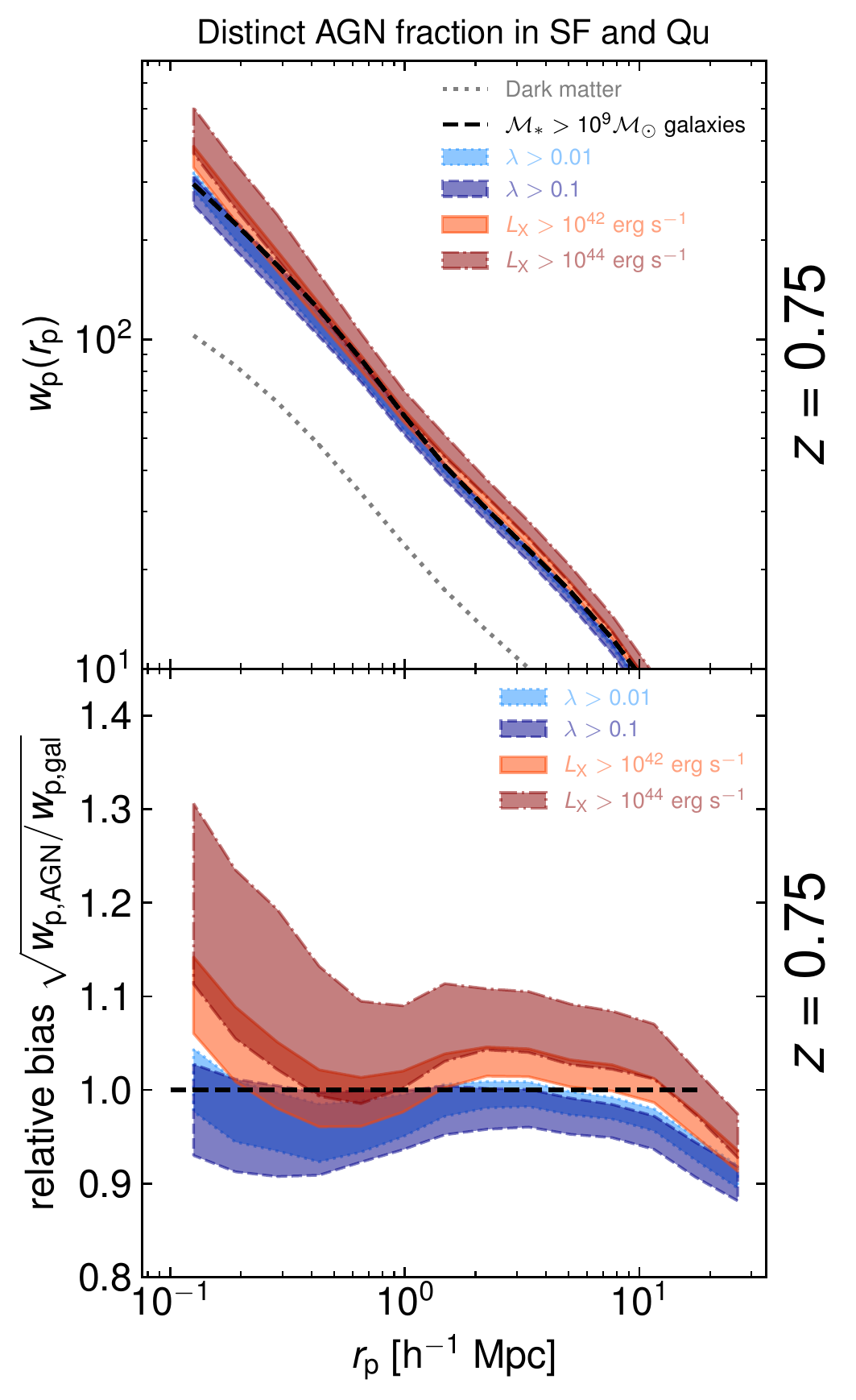}
    \caption{
    \textbf{Top:} Measurements of $w_p(r_p)$, comparing \LX-limited and $\lambda$-limited samples.
    \textbf{Bottom:} Relative bias as a function of $r_p$, relative to the parent $\mstel>10^{9}\msun$ galaxy sample, for the four AGN samples.
    The clustering amplitude of \LX-limited AGN samples are systematically higher than for $\lambda$-limited samples that account for observational biases.
    }
    \label{fig:wp_foursamples}
\end{figure}

\begin{figure}
    \centering
     \includegraphics[width=0.85\columnwidth]{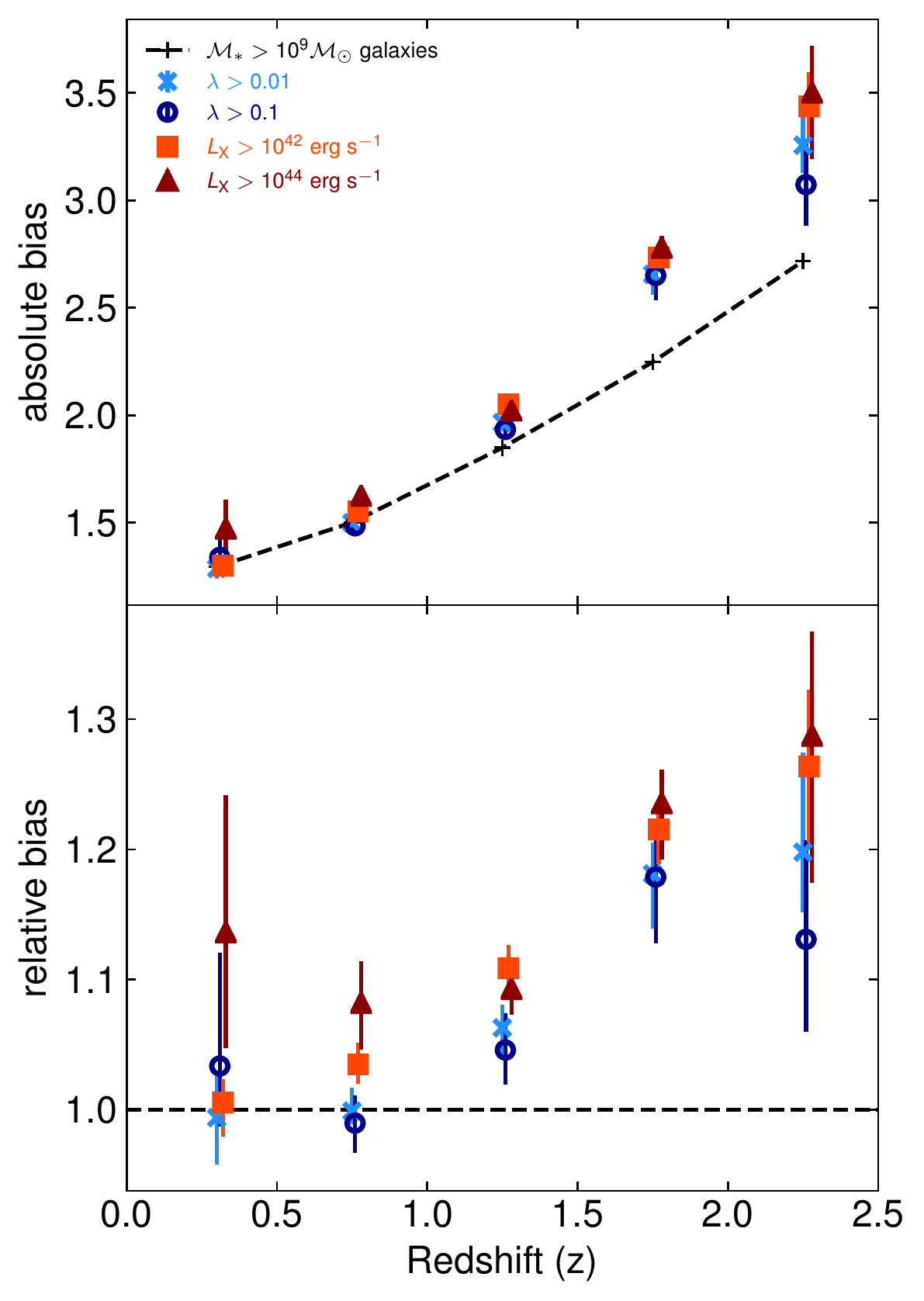}
    \caption{Absolute bias (relative to dark matter, top panel) and relative bias (relative to $\mstel>10^{9}\msun$ galaxies, bottom panel) as a function of redshift, averaged over $1 h^{-1}$~Mpc $< r_p < 10 h^{-1}$~Mpc, for four ``AGN samples" defined to different \LX\ and $\lambda$ limits, using our second (preferred) method to allocate AGN independently to star-forming and quiescent galaxies (as a function of \mstel). 
    The absolute bias of the AGN samples increases to higher redshifts, predominantly driven by the fact that \emph{galaxies} (black dashed line) are more strongly clustered relative to the underlying dark matter. 
    For the \LX-limited samples the bias predicted by our modelling lies systematically above the $\lambda$-limited samples. 
    The relative bias of $\lambda$-limited samples also increases to higher redshift due to the increased prominence of star-forming galaxies and the stronger \mstel\ dependence of the AGN fraction in this population.
    }
    \label{fig:bias_vs_z}
\end{figure}

In Figure~\ref{fig:wp_foursamples} we compare the predicted clustering properties for these four AGN samples at $z=0.75$. 
The differences in $w_p(r_p)$ are relatively small but we are able to identify clear systematic differences, which are seen most clearly in the relative bias compared to the clustering of galaxies (bottom panel).
The clustering of \sar-limited AGN samples is consistent with the clustering of the parent $\mstel>10^9\msun$ galaxy population, whereas the introduction of an \LX-limit (corresponding to a observational selection effect) leads to a increase in the bias, of up to $\sim10$\% at scales of $\gtrsim1h^{-1}$~Mpc.

In Figure~\ref{fig:bias_vs_z} we compare both the relative bias and the absolute bias (i.e. bias relative to the  underlying dark matter distribution) as a function of redshift for the four AGN samples and the underlying parent galaxy sample, where all bias measurements are averaged over scales of  \mbox{$1 h^{-1}$~Mpc $< r_p < 10 h^{-1}$~Mpc}.
We see a strong increase in the absolute bias toward higher redshift for all four AGN samples. 
Such an increase is expected and also seen in the parent galaxy population: at higher redshift, galaxies (of a given \mstel) and the dark matter haloes that they lie in correspond to more extreme over-densities relative to the underlying dark matter distribution.  
However, the {\it relative} bias of all four AGN samples, compared to galaxies, is also seen to increase with redshift (bottom panel). 
At $z<1$, the clustering properties of the \sar-limited AGN samples are consistent with the parent galaxy population; at higher redshifts these AGN samples are more strongly clustered, with the relative bias increasing to $\sim1.2$ (for $\sar>0.01$ AGN) at $z>2$, although we note the larger uncertainties at higher $z$ (due to the larger uncertainties in measurements of AGN fractions for higher redshift galaxies that are propagated into our model predictions).
The \LX-limited samples generally have a higher bias than the \sar-limited samples, due to the observational selection effects in such samples, although given the large uncertainties we only find a significant difference for our two best-sampled redshift bins (at $z=0.75$ and $z=1.25$).

Overall, at higher redshifts AGN activity shows a slight preference toward higher halo masses and thus denser environments relative to galaxies.
Such evolution is due to changes in the \mstel-dependence of AGN activity, whereby the fraction of AGN in higher mass star-forming galaxies  ($\mstel\gtrsim10^{10.5}\msun$) increases more strongly with redshift than in lower mass galaxies (see e.g. figure~6 of \PaperII), in addition to the increased dominance of such star-forming galaxies compared to quiescent galaxies (at a given \mstel) toward higher redshift.
The prevalence of AGN activity increases toward higher redshift due to changes in the properties of the \emph{galaxies} they lie in (e.g. higher gas fractions leading to higher SFRs and a greater rate of AGN triggering). 
We do \emph{not} infer any direct connection between the large-scale environment and an increased incidence of AGN activity.

\section{Comparison to Observed AGN clustering measurements}
\label{sec:observations}

We next compare the clustering predictions of our model with observed AGN clustering measurements. We focus first on comparing to \LX-limited AGN samples that span a range of luminosity and redshift.  In Figure 11 in the left panel we present a variety of published bias measurements for \LX-limited samples from $z\sim0$ to $z\sim2$. The observed bias values generally increase with redshift, from $\sim1$ at $z\sim0$ to $\sim3$ at $z\sim2$.  The dominant  observational errors are cosmic variance and Poisson noise due to small AGN sample sizes. 
These studies span a range of luminosities, though they typically target AGN with \LX$\sim10^{43}$~\ergs.  We also show predictions for the absolute bias as a function of redshift for our preferred model, for two values of \LX.  In our preferred model the clustering amplitude is a mild function of the \LX\ limit of the AGN sample, rising only at the highest luminosities. As can be seen in the figure, the predictions from our model match the observational measurements of the absolute bias well, for a wide range of observational studies.

\begin{figure*}
            \includegraphics[height=6.8cm,trim=0 0 0 0]{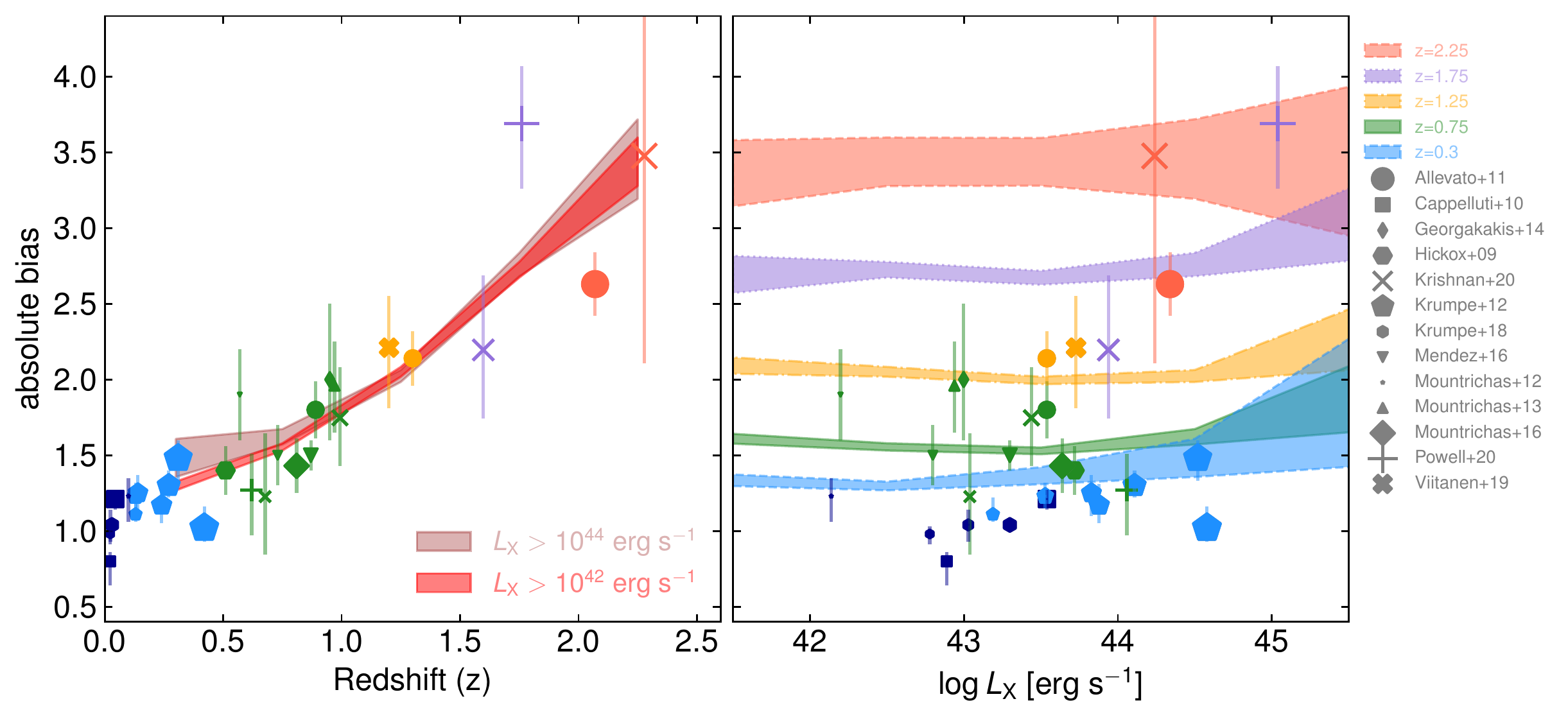}
    \caption{
    \refone{Our model predictions for the absolute bias as a function of redshift (left, coloured regions indicating two \LX-limited AGN samples) and as a function of X-ray luminosity (right, coloured regions indicating different redshifts), compared to observational measurements of the bias from X-ray selected samples. 
    Symbol types indicate the prior study and the sizes are scaled based on the average luminosity of the AGN in these samples. 
    The symbol colours associate the observational measurements to one of our redshift bins (measurements at $z<0.1$, below the range of our AGN fraction measurements, are indicated in dark blue). 
    Our predictions are consistent with direct measurements of X-ray AGN clustering and the overall increase in the absolute bias to higher redshift.
     Our model predicts little, if any, dependence of the absolute bias on X-ray luminosity, which is generally consistent with the findings of prior studies.}
    }
    \label{fig:obs_bias_vs_lx_and_z}
\end{figure*}
\nocite{Allevato11,Cappelluti10,Georgakakis14b,Hickox09,Krishnan20,Krumpe12,Krumpe18,Mendez16,Mountrichas12,Mountrichas13,Mountrichas16,Powell20,Viitanen19}

\begin{figure}
    \includegraphics[height=6.8cm]{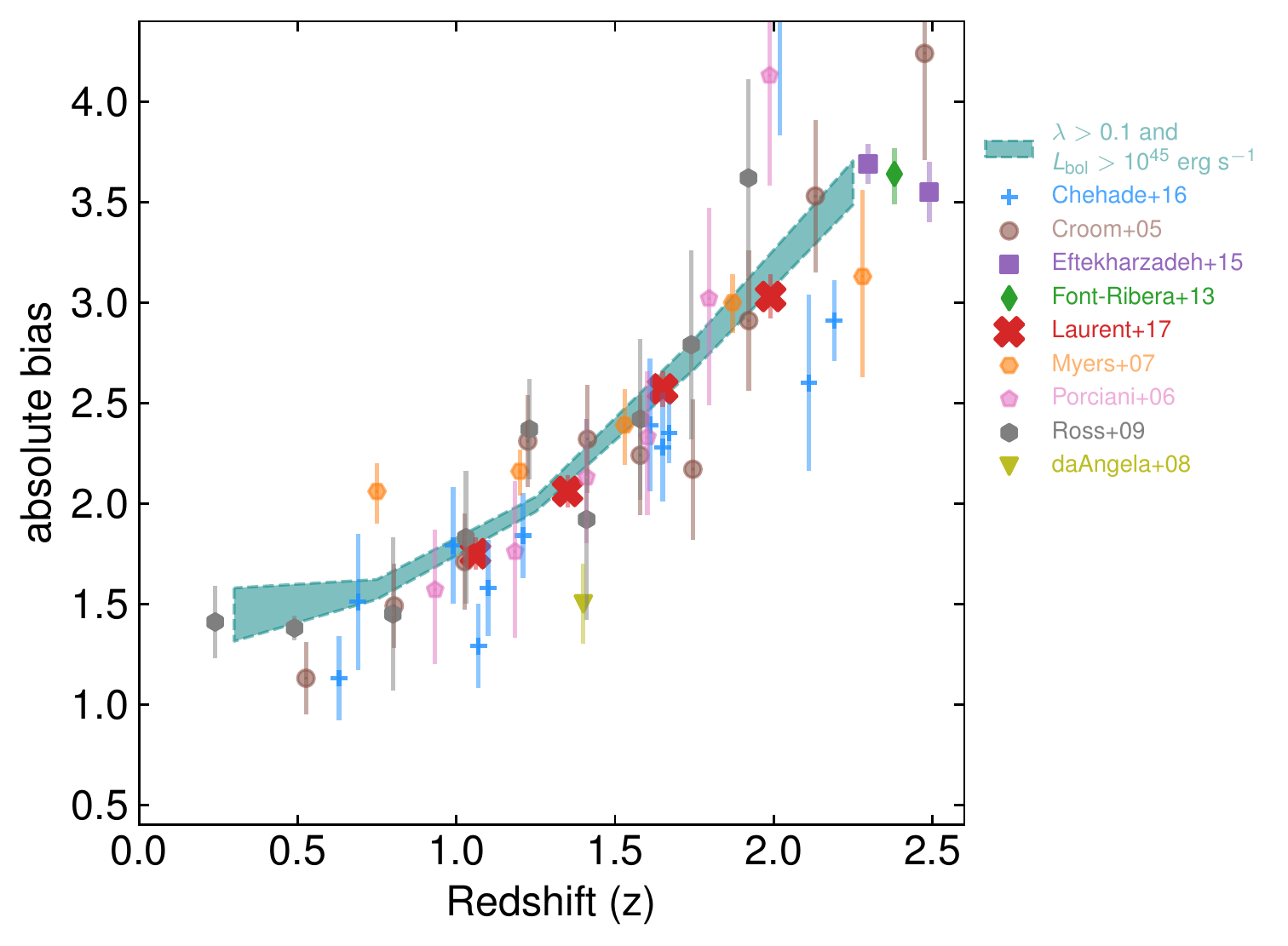}
    \caption{
    Our model prediction for the absolute bias as a function of redshift for a ``quasar sample", which we approximate with cuts of both $\lambda>0.1$ and bolometric luminosity $L_\mathrm{bol}>10^{45}$~\ergs (teal region), compared to  measurements of the bias for optically-selected quasar samples (symbols and colours indicating the different studies, as shown). 
        Our model prediction is generally consistent with observed quasar samples out to $z\sim2.5$.   
    \refone{We do not scale the symbol sizes to indicate differences in the luminosity limits of the observed quasar samples, in contrast to Figure~\ref{fig:obs_bias_vs_lx_and_z} for X-ray selected samples, although we note that none of the studies find a strong luminosity dependence at a fixed redshift, consistent with our model.} 
    }
    \label{fig:quasar_bias_vs_z}
\end{figure}
\nocite{Chehade16,Croom05,Eftekharzadeh15,Font-Ribera13,Laurent17,Myers07,Porciani06,Ross09,daAngela08}

In the right panel of Figure~\ref{fig:obs_bias_vs_lx_and_z} we show the same absolute bias values from the literature, here displayed as a function of \LX.  To highlight the differences in the redshifts of each study, we colour code the data points from blue at the lowest redshifts to red at the highest redshifts.  The coloured regions show our model predictions at different redshifts, as a function of luminosity.  As discussed above, our model has little, if any, luminosity dependence, and as can be seen it  agrees well with the lack of luminosity-dependence in the observational results.  Our model does not extend to $z\sim0$, where the observational data are shown in dark blue.  We note in particular the good agreement with the $z=1.5 - 2.0$ results (from \citealt{Krishnan20} and \citealt{Powell20}, shown in purple in the right panel), which do show a substantial rise in the absolute bias from \LX$\sim10^{44} - 10^{45}$~\ergs.
To produce such high luminosities requires massive galaxies (and thus massive black holes) with high accretion rates. Such sources are thus restricted to the most extreme and massive haloes, especially at high-$z$, leading to our predicted rise in the absolute bias, consistent with the observed data.

We compare in Figure \ref{fig:quasar_bias_vs_z} the predictions from our model to clustering measurements of optically-selected quasar samples. 
Such quasar samples likely suffer from a combination of selection biases, selecting high-luminosity, unobscured quasars whose light dominates that of their host galaxies (i.e. high-$\lambda$).
In order to compare with these observational samples, we restrict our model predictions in this figure to both a high luminosity {\it and} high $\lambda$ AGN sample, with a bolometric luminosity of $L_\mathrm{bol}>10^{45}$~\ergs and $\lambda>0.1$.  This combination of a luminosity cut and a $\lambda$ limit brings our model into excellent agreement with the observational results and in particular the recent high-precision measurements from the SDSS-IV eBOSS quasar samples at $z>1.5$ \citep{Laurent17}.  We note that---given the uncertainties in our model and the mild dependence of the bias on either $\lambda$ of \LX\ (e.g. Figure~\ref{fig:bias_vs_z})---a simple, high luminosity cut can also provide good agreement with the observed data. 

It has been noted in the literature that quasar clustering results are surprisingly consistent with quasars residing in halos of roughly constant mass across the bulk of cosmic time \citep[e.g.,][]{Porciani04, Eftekharzadeh15}.  We show here that this is a natural outcome of the combination of the selection biases inherent to optical quasar samples and the known AGN occupation of galaxies.  We also show that the halo mass distribution is wide, and that quasars (and all AGN) reside in a range of halo masses at all redshifts.  The width of this distribution is not obvious from a simple statistic like the bias. 

As noted above, 
our model predicts small ($\sim10$\%) systematic differences between \LX-limited and $\lambda$-limited AGN samples.
As shown in Figure 10, at intermediate redshift \LX-selected AGN samples have slightly higher bias than $\lambda$-selected AGN samples.
Observationally, these are similar to typical \LX-limited and optical quasar samples. Our model predictions could therefore explain some of the indications in the literature that X-ray selected AGN samples have slightly elevated clustering compared to optical quasar samples \citep[e.g.,][]{Coil09}.  However,  to test this observationally one should use AGN samples identified from the same parent galaxy survey, with overlap in the optical and X-ray selected AGN samples, and any differences in luminosity between the samples should be taken into account \citep[e.g.,][]{Krumpe12}.  
Regardless, the differences between the clustering amplitudes of optical quasar and X-ray selected AGN samples should {\it not} be interpreted as different stages of AGN evolution \citep[e.g.,][]{Hickox09} but rather as due to differences in the selection biases of AGN identified in different ways.

\begin{figure*}
\centering
\includegraphics[height=7.2cm,trim=0.0cm 0.3cm 0cm 0]{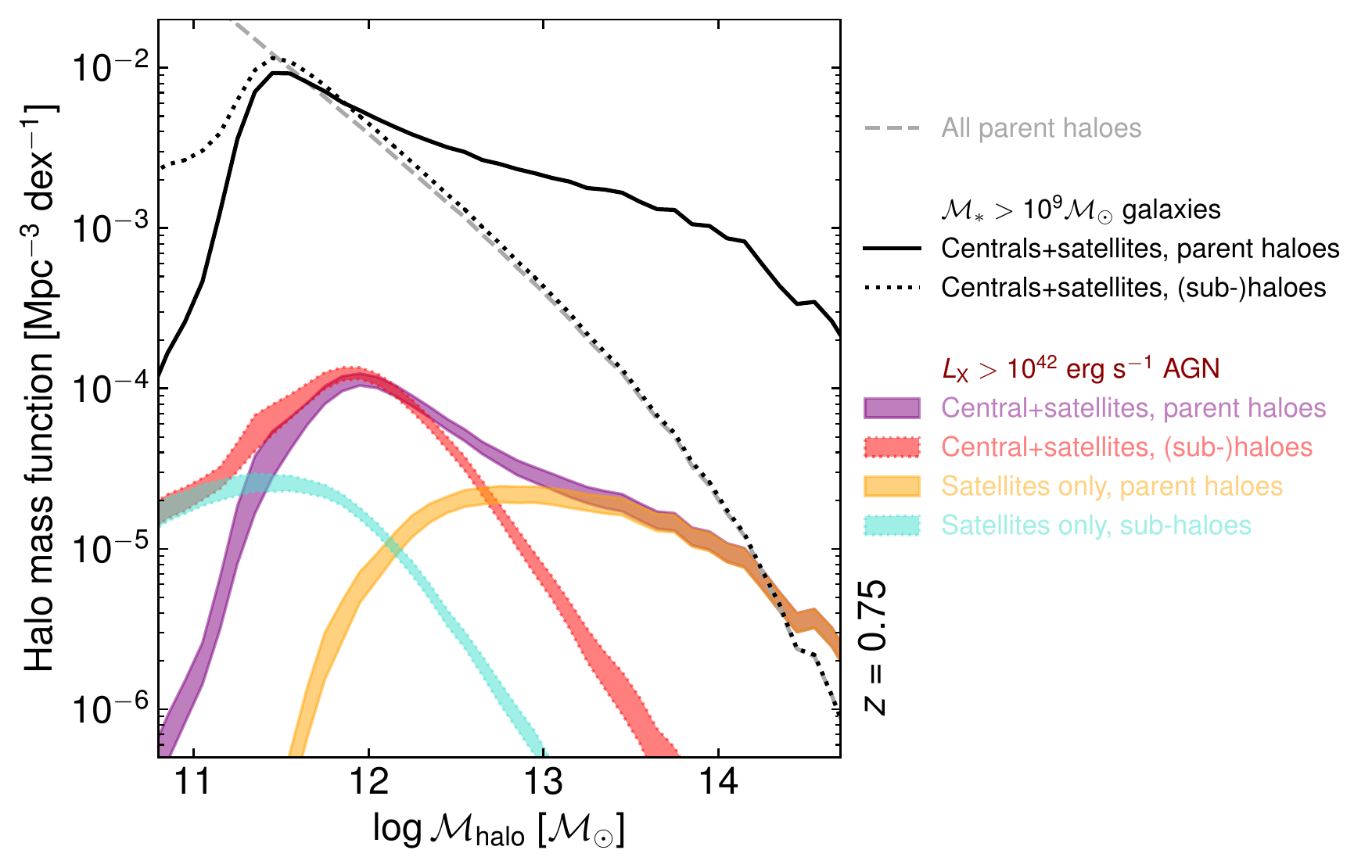}
\caption{
Comparison of predicted HMFs at $z=0.75$ for different definitions of the halo mass. 
The grey dashed line shows the mass function of all individual parent haloes in the simulation.
The solid black line shows the parent-halo mass function for all $\mstel>10^{9}\msun$ galaxies (adopting the current virial mass of the halo); high-mass parent haloes contain multiple galaxies (both centrals and satellites) and are thus counted multiple times for this HMF, resulting in the broad flat shape. 
The black dotted line shows the halo mass function when adopting the current virial (sub-)halo mass (the parent mass for central galaxies and the sub-halo mass for satellites), which has a much steeper shape at high masses and a tail extending to lower mass.
The purple region shows the HMF of parent haloes for $\LX>10^{42}$~\ergs\ AGN (in both centrals and satellites), whereas the red region adopts the current \subhalo\ mass.
The yellow and cyan regions show the very different HMFs when considering the parent and sub-halo mass, respectively, for only the AGN that lie in satellite galaxies.}
\label{fig:hmf_definitions}
\end{figure*}
\begin{figure*}
\centering
\includegraphics[height=7.2cm,trim=0cm 0.3cm 0cm 0]{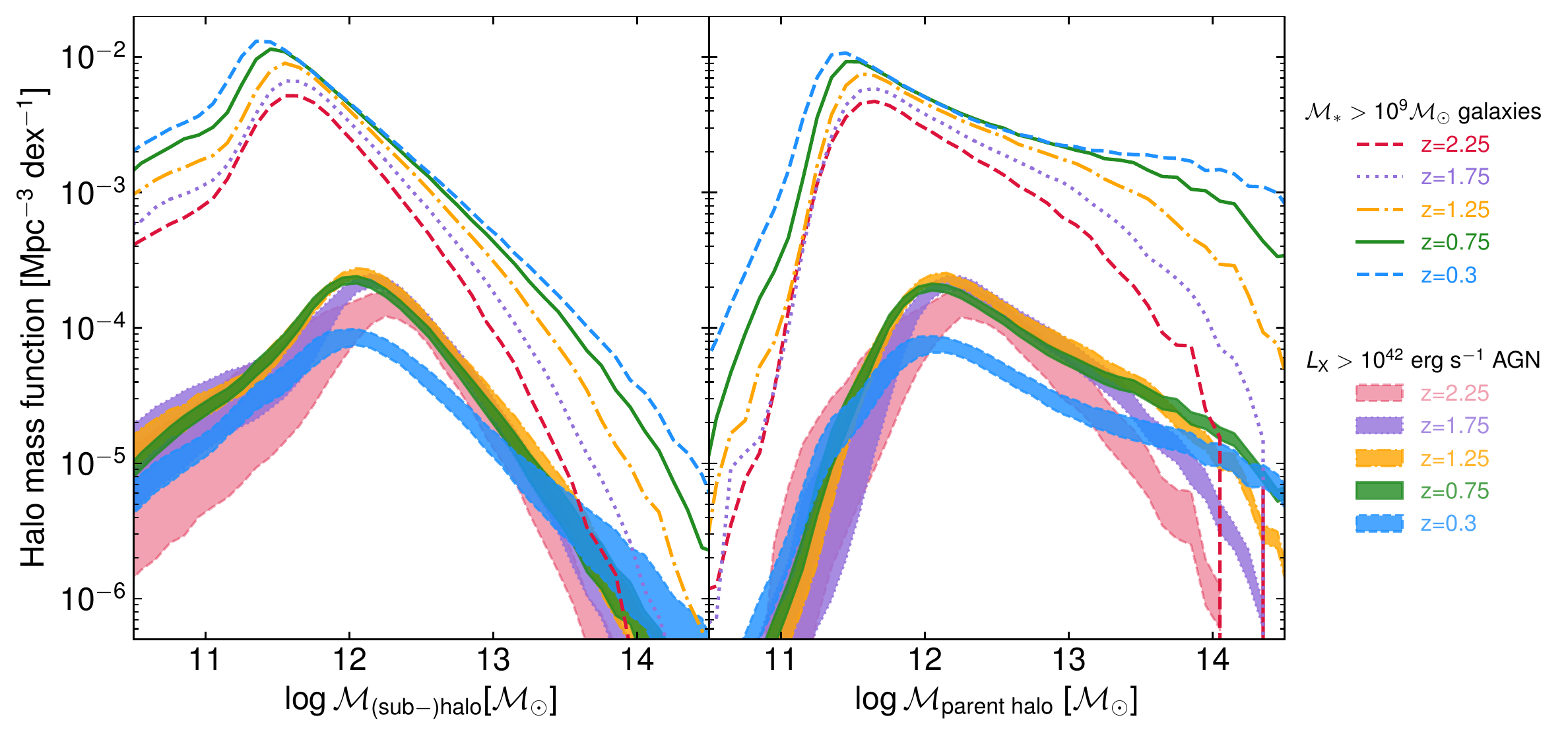}
\caption{
HMFs of galaxies (lines) and $\LX >10^{42}$~\ergs\ AGN (shaded regions) as a function of redshift (as indicated by colours) for the \subhalo\ mass (left) and parent halo mass (right).}
\label{fig:hmf_vs_redshift}
\end{figure*}

\section{Discussion and interpretation}

Having established our forward model approach to populate haloes with AGN, based on measurements of AGN fractions as a function of galaxy properties, 
and shown that the clustering properties predicted by this model are in good agreement with previous studies,
we now move on to discuss some of the implications of our model.
In Section~\ref{sec:parent_vs_sub} we discuss the properties of the haloes that host AGN, in particular focussing on the important distinction between the \subhalo\ containing an individual galaxy (and its AGN) and the larger parent dark matter halo that 
defines the broader scale environment. 
In Section~\ref{sec:interpret_bias} we show how commonly-used methods that convert measurements of the absolute bias to a typical halo mass provide an inaccurate and incomplete picture of the hosts of AGN, neglecting the underlying distribution of halo masses, the AGN satellite fraction, and AGN selection effects. 
In Section~\ref{sec:hod} we use our model to predict the halo occupation distribution (HOD) of AGN,
present predictions of the AGN satellite fraction, and discuss the advantages and limitations of HOD modelling in the interpretation of AGN clustering measurements.
Finally, we discuss the importance of interpreting AGN clustering measurements compared to the underlying galaxy population in Section~\ref{sec:galaxybias}.
.

\subsection{Parent and sub-halo mass distributions}
\label{sec:parent_vs_sub}

In the results presented in Section~\ref{sec:agn_hmfs_and_wp} above we adopted the peak historical (sub-)halo mass, \mpeak, as our tracer of the halo mass when presenting the HMFs of AGN. 
\mpeak\ traces the overall growth of the \subhalo\ in which a galaxy forms and is thus closely related to the overall stellar build-up of a galaxy (i.e. the SFR and total \mstel).
 The triggering of AGN activity requires gas to be driven into the very central region of a galaxy and it is thus the small-scale environment---i.e. the host galaxy properties---that should determine the incidence of AGN; this assumption is inherent in our model.
 However, it is the mass of the larger parent halo that an AGN host galaxy lies in, \mparent, that traces the large-scale environment and thus determines the \emph{clustering} properties of AGN samples.

In Figure~\ref{fig:hmf_definitions} we present HMFs using different definitions of the halo mass. 
Massive parent haloes contain both a central galaxy and a large number of satellite galaxies. As such, a single parent halo is counted multiple times when constructing the {\it parent} halo mass function of all galaxies above a chosen \mstel\ threshold (solid black line), producing a relatively flat distribution extending out to $\mparent\sim10^{14}\msun$. 
When the current virial \subhalo\ mass (i.e.~the parent halo mass for central galaxies but the mass of the individual sub-halo of a satellite galaxy) is adopted instead, the HMF has a much steeper high-mass slope and a tail at masses below $\msub\sim3\times10^{11}\msun$ (black dotted line). 

AGN samples exhibit a similar behaviour; our prediction for $\LX > 10^{42}$~\ergs\ is shown in Figure~\ref{fig:hmf_definitions}.
The parent halo mass function, including AGN in both centrals and satellites, has a broad tail extending to high masses (purple region), whereas the mass function for the individual \subhaloes\ of AGN extends to lower masses (red region), due to differences in the parent halo masses (gold region) and sub-halo masses (cyan region) of AGN in satellite galaxies. 
{\it The typical parent halo mass of AGN is thus significantly higher than the typical \subhalo\ mass}: 
averaging over the HMFs \refone{to calculate the mean halo mass} gives $\langle \mparent \rangle \approx 10^{13}\msun$ versus 
$\langle \msub \rangle\approx 2\times 10^{12} \msun$ for $\LX>10^{42}$~\ergs\ AGN at $z=0.75$.
The difference in these masses reflects the fact that AGN are often found in satellite galaxies that lie within massive parent haloes \citep[see Section~\ref{sec:hod} below, see also][]{Alam20}.

In Figure~\ref{fig:hmf_vs_redshift} we explore how the parent and \subhalo\ mass functions evolve with redshift. 
The normalization of the galaxy HMF (coloured lines) increases at all masses as cosmic time progresses (i.e.~as redshift decreases), reflecting the ongoing hierarchical build up of dark matter structure. 
The rate of increase is especially strong for parent halo mass functions at higher masses ($\mparent \gtrsim 10^{13}\msun$), which is due to the build up of the satellite galaxy population as smaller haloes fall into larger parent haloes. 
The most massive parent haloes, $\mparent\gtrsim10^{14}\msun$, are yet to form at $z\gtrsim2$, leading to 
an even stronger evolution as this population grows. 

The HMFs of AGN follow a different evolutionary pattern. 
The \subhalo\ mass function (shaded regions in the left panel of Figure~\ref{fig:hmf_vs_redshift}) remains roughly constant in both normalisation and shape between $z\sim2$ and $z\sim0.75$. 
At this epoch, AGN activity is found across the galaxy population and over a wide range of halo masses, albeit with an increased triggering rate in higher stellar mass galaxies (an effect that is further exaggerated by observational selection effects). 
At lower redshift, the overall normalisation of the \subhalo\ mass function drops by a factor $\sim$3.
While the large-scale structure of the Universe (and the galaxies that lie within it) builds up progressively with time, AGN are short-lived and the rate at which they are triggered drops significantly at later cosmic times. 
The peak of the \subhalo\ mass function also shifts toward slightly lower values as the \mstel-dependence of the AGN fraction becomes weaker at lower redshifts, as well as becoming broader due to the build-up of the highest mass haloes.
The parent halo mass function of AGN (shaded regions in the right panel of Figure~\ref{fig:hmf_vs_redshift}) follows broadly similar patterns, except at the highest masses ($\mparent\gtrsim10^{13}\msun$) where the build up of the most massive parent haloes---in particular by the accretion of satellite galaxies that often host AGN---leads to an overall increase.

Regardless of which halo mass definition is used, it is clear that AGN are found in haloes with a wide range of masses.
While clustering measurements broadly reflect the underlying distribution of parent halo masses, 
to understand the connections between AGN and galaxy evolution it is vital to understand the sub-haloes that the host galaxies and their AGN reside in.

\subsection{Interpretation of AGN bias measurements}
\label{sec:interpret_bias}

Our forward modelling approach allows us to predict the absolute bias that would be measured for different AGN samples using the standard observational approach: measuring the projected two-point correlation function of AGN, $w_p(r_p)$, and comparing to the corresponding correlation function of dark matter at scales of $r_p\gtrsim1h^{-1}$~Mpc. 
In Section~\ref{sec:observations} we showed that our predicted bias measurements are in good agreement---after applying appropriate observational limits---with a wide range of results for both X-ray AGN samples and optically-selected quasar samples 
out to $z\sim2.5$. 
Many prior studies then use their measurements of the absolute bias to infer the ``typical" halo mass of AGN using a simplistic comparison to dark matter simulations
\citep[e.g.][]{Croom05,Cappelluti10,Georgakakis14b}.
Here, we show explicitly that using a measured bias to directly infer a halo mass, 
without taking into account the satellite fraction or underlying mass distribution,
leads to estimates that are systematically offset from the true averages of {\it either} the \subhalo\ or parent halo masses \citep[see also][]{DeGraf17,Powell20}.

\begin{figure}
    \centering
    \includegraphics[width=0.90\columnwidth]{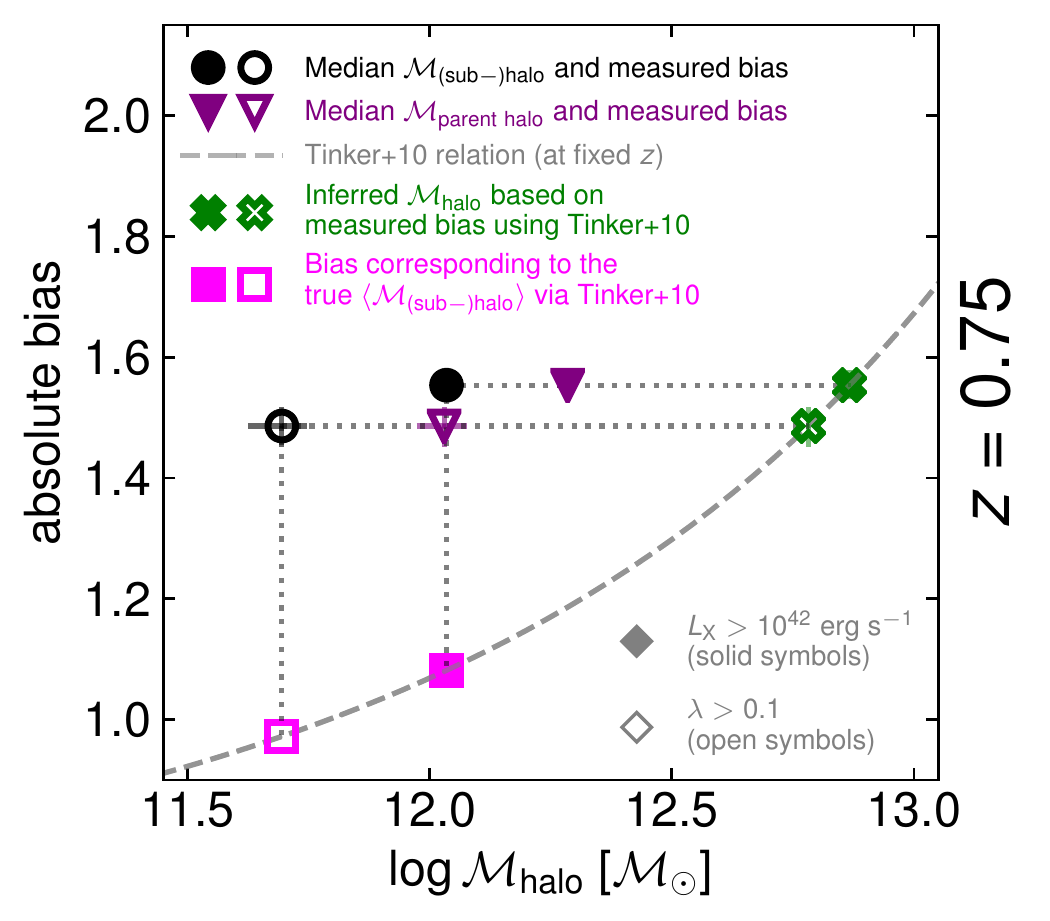}
    \caption{
    Demonstration of how measurements of the absolute bias may be used to (erroneously) infer the typical halo mass of AGN samples. 
   \refone{The solid black circle indicates our model prediction of the measured bias for an $\LX>10^{42}$~\ergs\ AGN sample and the median \msub\ based on our halo mass functions (open symbols show the equivalent measurements for a $\lambda>0.1$ AGN sample).} The purple triangles indicate the \refone{median} \mparent\ for the same AGN samples.
    The grey dotted line shows the relationship between bias and halo mass derived by \citet{Tinker10} \refone{using narrow bins in halo mass.}
    This relationship is often used to infer a ``typical" halo mass (green cross) for an AGN sample that corresponds to neither the true (sub-)halo mass nor the true parent halo mass. 
    The pink squares shows how we can instead determine a bias corresponding to the \refone{true median} $\msub$ using the \citet{Tinker10} relation. 
    }
    \label{fig:bias_demo}
    
\end{figure}

To convert a measured absolute bias into a typical halo mass, a common approach in AGN clustering studies uses the relationship between halo mass and bias derived from N-body simulations \citep[e.g.~][]{Sheth01,Tinker10,Comparat17}. 
Results from simulations estimate the bias for a \emph{narrow} distribution of halo masses, which differs substantially from the broad distributions of halo masses for AGN samples that are predicted by our model. 
More importantly, such studies implicitly assume one galaxy per halo, i.e.~only central galaxies are considered.
In Figure~\ref{fig:bias_demo} we show the \citet{Tinker10} relationship between \mhalo\ and absolute bias, evaluated at $z=0.75$,
which we compare to the \refone{median} \subhalo\ and parent halo masses predicted by our model (black circles and purple triangles, respectively) and the corresponding absolute bias that we measure in a manner analogous to observational studies (i.e.~via projected two-point correlation functions). 
The median halo masses can vary by {\it up to 0.5~dex} depending on the selection method (\LX-limited or $\lambda$-limited) and whether parent or \subhalo\ mass is considered. 
The green crosses show the estimates of \mhalo\ that would be inferred using the \citet{Tinker10} relation, which 
\refone{are significantly above the median \msub\ and \mparent.}
Using such a relation to infer a halo mass from the absolute bias 
ignores the broad distribution of halo masses and does not account for the fraction of AGN that lie in satellite galaxies within massive parent haloes. 
\refone{We also note that \emph{mean} halo masses are significantly skewed by the high-mass tail of the HMF, especially in the case of the mean AGN parent halo mass. 
We choose to report median halo masses as these are more representative of the bulk of AGN and are robust against uncertainties in the shape of the HMF for the most massive host haloes. 
We note that none of the mean, median or mode halo masses, for parent haloes or \subhaloes, are in agreement with the ``typical'' halo mass inferred from the bias using the \citet{Tinker10} relation.
} 

\begin{figure*}
\centering
\includegraphics[width=0.6\textwidth,trim=1.0cm 0.3cm 1.1cm 0.2cm]{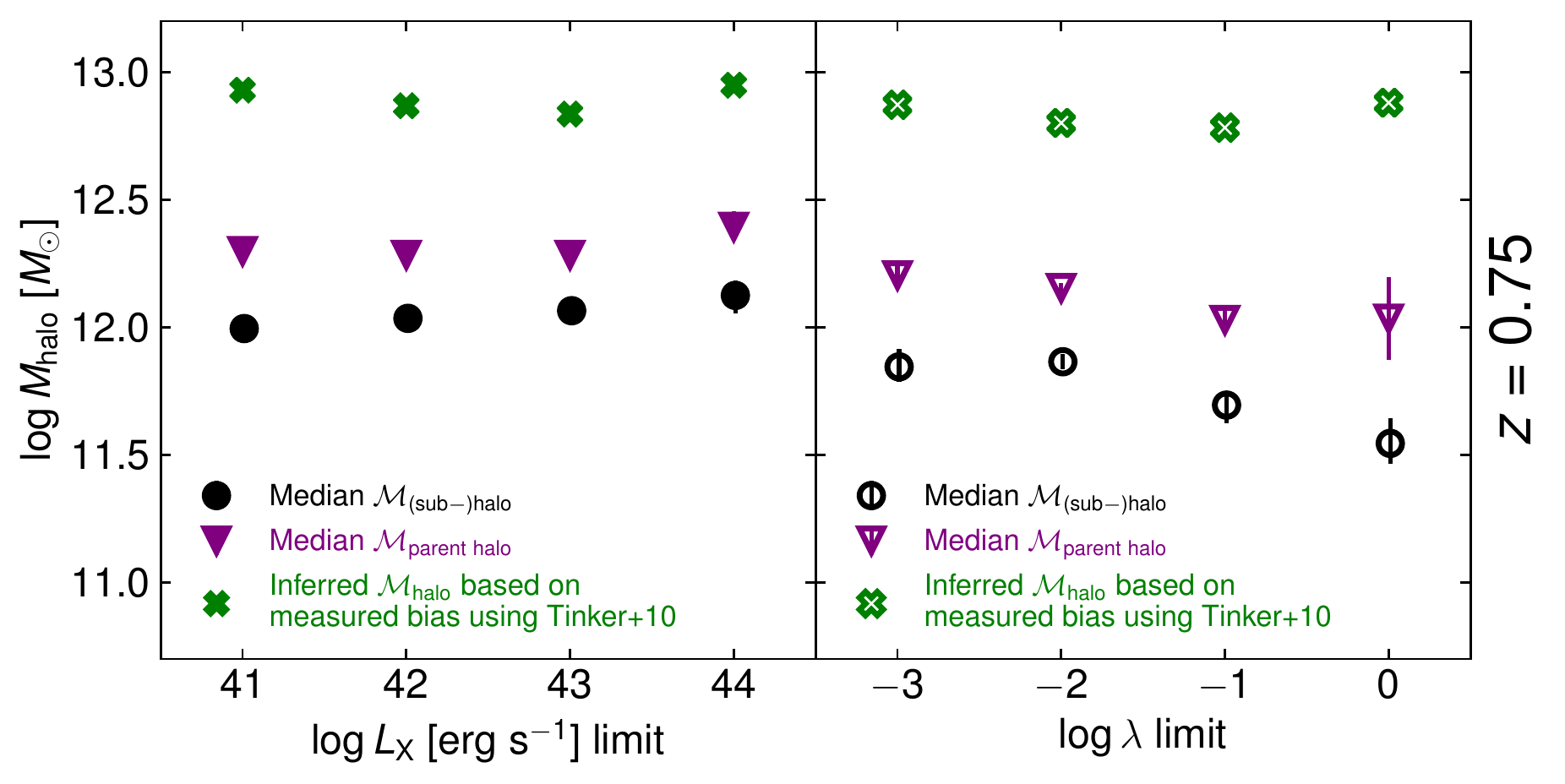}
\caption{
Halo mass as a function of \LX\ limits (left) and $\lambda$ limits (right) at $z=0.75$, showing the true \refone{median} mass of (sub-)haloes (\refone{black} circles) and parent haloes (purple triangles) from our model compared to the halo mass that would be inferred based on the measured bias using the \citet{Tinker10} relation (green crosses). }
\label{fig:avMhalo_vs_lims}
\end{figure*}
\begin{figure*}
\includegraphics[width=0.6\textwidth,trim=1.0cm 0.3cm 1.1cm 0.5cm]{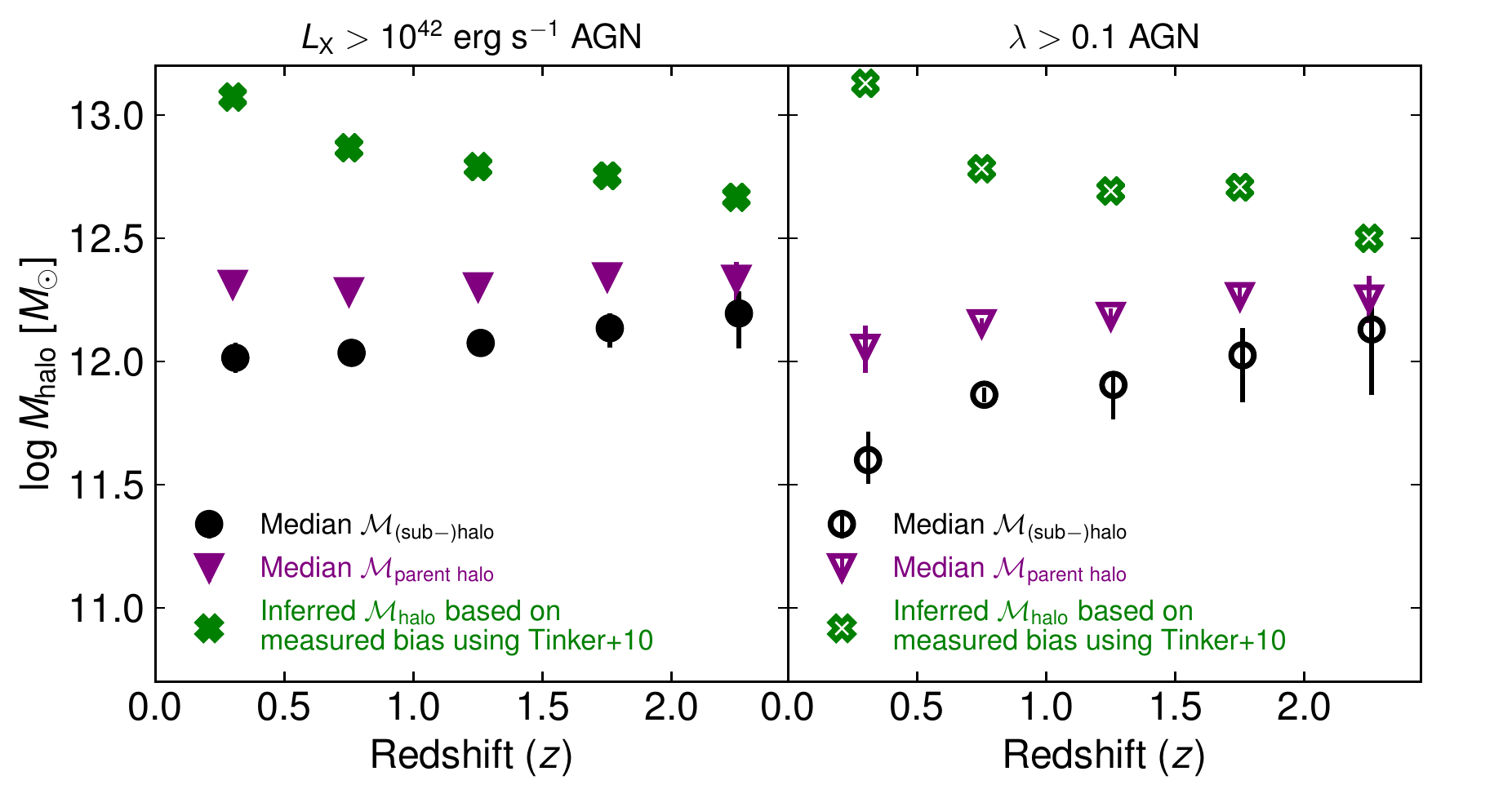}
\caption{
Halo mass as a function of redshift, showing results for the
$\LX>10^{42}$~erg~s$^{-1}$ (left panel) and $\lambda>0.1$ (right panel) samples.
The \refone{median} parent halo mass of AGN samples (purple triangles) is found to decrease with increasing redshift, whereas the mean \subhalo\ mass (\refone{black} circles) remains relatively constant. Inferring \mhalo\ directly from the bias \citep[e.g.~using][; green crosses]{Tinker10} provides an inaccurate estimate of AGN halo masses. 
}
\label{fig:avMhalo_vs_redshift}
\end{figure*}

\begin{figure*}
\centering
\includegraphics[width=0.46\textwidth,trim=2.5cm 0.5cm 0cm 0]{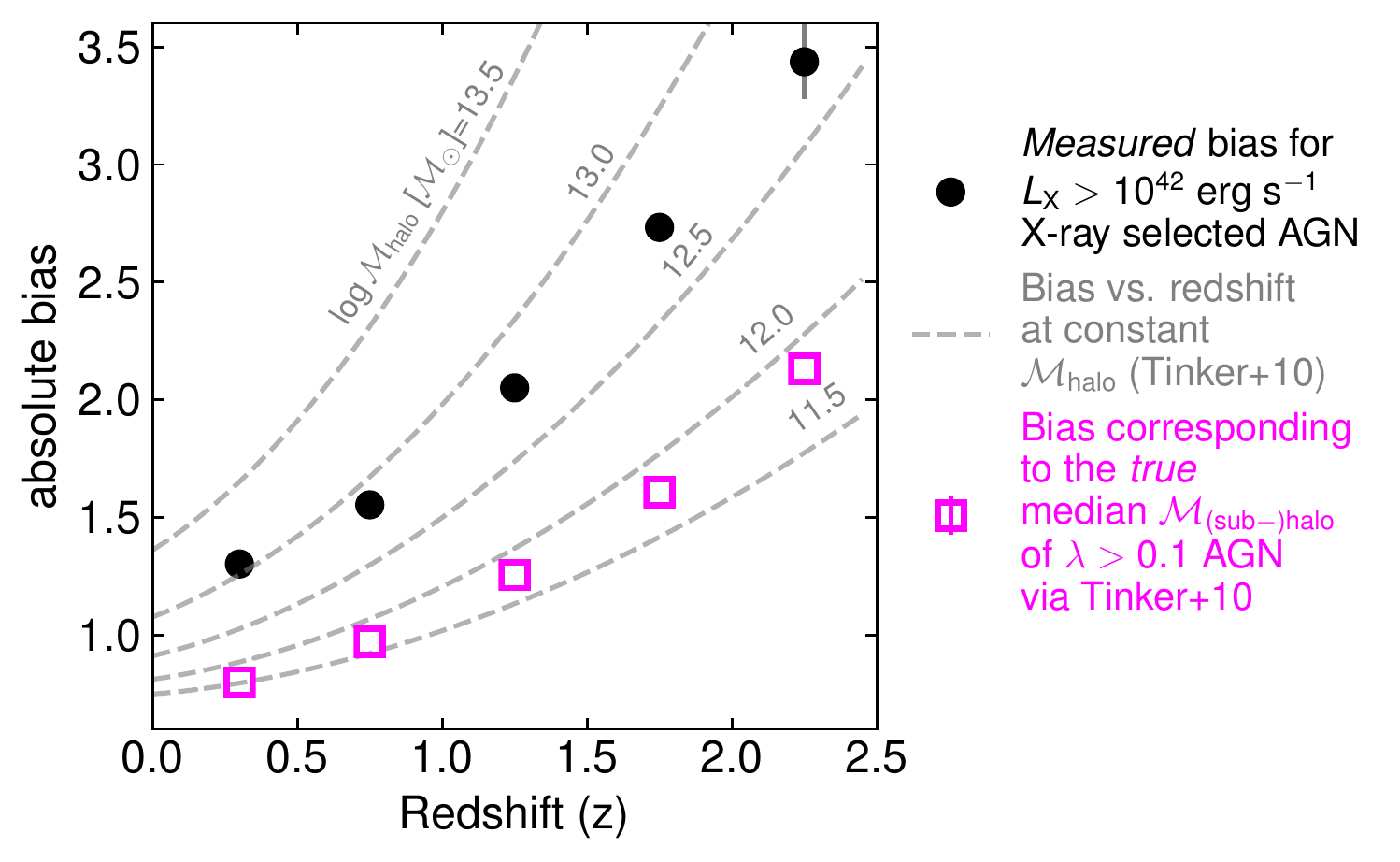}
\
\caption{
Absolute bias as a function of redshift, showing our predictions of the measured bias for $\LX>10^{42}$~\ergs\ AGN (black circles) compared to the bias of dark matter haloes of constant mass \citep[grey dashed lines:][]{Tinker10}. 
Such relations are often used to interpret the measured bias of AGN samples and  (erroneously) conclude that AGN all lie in dark matter haloes with $\log \mhalo \approx 12.5-13.0$. 
The pink open squares show the bias that corresponds to the true \refone{median} $\msub$ for a 
$\lambda$-limited AGN sample.
Using the measured bias of AGN samples to infer the typical halo mass 
using relations derived from simulations with narrow halo mass bins \citep[e.g.][]{Tinker10,Comparat17}
leads to an overestimate of $\sim0.5-1$~dex.
This widely used methodology neglects the distinction between parent haloes and sub-haloes, does not account for known observational selection biases in AGN samples, and does not capture the true breadth of the distribution of AGN host halo masses.
}
\label{fig:bias_vs_z_constmhalo}
\end{figure*}

\refone{In Figure~\ref{fig:avMhalo_vs_lims} we show how these measures of halo mass---the median \msub\ and \mparent\ as well as the \mhalo\ that is inferred using the \citet{Tinker10} relation---vary depending on the chosen \LX\ (left) or $\lambda$ (right) limit, shown at $z=0.75$.
We find little dependence on the chosen \LX\ limits (at a fixed redshift), reflecting the relatively minor changes in the shapes of the underlying HMFs (see Section~\ref{sec:agn_hmfs_and_wp}), although we note that subtle (but significant) changes in the shape are not captured in a single statistic such as the median.
In contrast, we find a slight \emph{decrease} in the median \subhalo\ masses with increasing $\lambda$ limit, which is related to the weaker dependence on stellar mass of the AGN fraction for high-$\lambda$ samples (see Figure~\ref{fig:agnfractions} and discussion in Section~\ref{sec:agnfrac}) that brings the AGN HMFs closer to the HMF of the underlying galaxy population (see Figure~\ref{fig:hmfs_lambdalims}).
}

Figure~\ref{fig:avMhalo_vs_redshift} shows how our different halo mass estimates vary as a function of redshift for the  $\LX>10^{42}$~\ergs\  (left) and $\lambda>0.1$ (right) AGN samples.
\refone{The median \subhalo\ masses (black circles), are found 
to be roughly constant with redshift for both samples,
reflecting the lack of evolution in the AGN halo mass functions with redshift (see Figure~\ref{fig:hmf_vs_redshift}).
However, there is evidence of a decrease at the lowest redshift for the $\lambda>0.1$ sample that corresponds to the reduced stellar mass dependence of the AGN fraction for such samples at the lowest redshifts.}
The immediate environment in which AGN activity is triggered, traced by the mass of the host \subhaloes, does not appear to change substantially with cosmic time.
However, we note that AGN are found over a broad range of \subhalo\ masses and there \emph{are} important changes in the triggering rate depending on the properties of the host \emph{galaxies}, which will be closely related to the immediate environment of the supermassive black holes and the efficacy of different physical processes to bring gas into the central regions to trigger AGN activity. 

\refone{The median parent halo mass of AGN samples (purple triangles in Figure~\ref{fig:hmf_vs_redshift}) are $\sim$0.1--0.3~dex higher than the corresponding median \subhalo\ masses and remain approximately constant at all redshifts ($\sim2\times10^{12}\msun$ for $\LX>10^{42}$~\ergs\ AGN).
Thus, half of the AGN in \LX-limited samples are found in haloes with $\mparent\lesssim 2 \times 10^{12}\msun$ at $z<2.5$.}
\refone{We note that the median \mparent\ is not affected by the build up of the most extreme, massive parent haloes.
In contrast, the \emph{mean} parent halo mass of AGN rises by almost an order of magnitude as cosmic time progresses, increasing from $\sim10^{12.6}\msun$ at $z\sim2.5$ to $\sim10^{13.5}\msun$ at $z\sim0.3$ for $\LX>10^{42}$~\ergs\ AGN. 
As time progresses, the most massive haloes start to assemble and accrete smaller sub-haloes, containing satellite galaxies where AGN activity often occurs (see Figure~\ref{fig:hmf_vs_redshift}, right). } 
\refone{It is the build up of these satellite galaxy populations within massive parent haloes that drives the increase in the mean $\mparent$ with cosmic time.}
We do not infer any direct correlation between the parent halo---and thus the large-scale environment---and AGN activity.

\refone{
The green crosses in Figure~\ref{fig:avMhalo_vs_redshift} lie significantly above the median \msub\ and \mparent\ and show a mild decline with increasing redshift, but do not accurately reflect the evolution of the mean \mparent. The satellite fraction and broad parent halo mass distribution are not accounted for when using the absolute bias to infer a halo mass and thus they do not provide an accurate tracer of the typical halo mass.}

To further illustrate the dangers of using the absolute bias to simplistically infer a halo mass, in Figure~\ref{fig:bias_vs_z_constmhalo} we show the measured bias for an observational, luminosity-limited sample of AGN (red points, as predicted by our model) as a function of redshift, compared to the absolute bias as a function of redshift given by \citet{Tinker10} for haloes of constant \mhalo and with a narrow distribution (grey dashed lines). 
Such plots are often used to summarise observational results and suggest that AGN activity occurs uniformly in haloes of mass $\sim 10^{12.5-13.0}\msun$ throughout cosmic time \citep[e.g.][]{Ross09,Chehade16,Mendez16}.
The pink squares instead show the absolute bias that corresponds to the median \msub\ for a complete $\lambda>0.1$  AGN sample 
i.e.~indicating the absolute bias that {\it would be measured} for a narrow distribution of haloes 
at the mass corresponding to the true median \msub\ 
\citep[as given by][see also Figure~\ref{fig:bias_demo}]{Tinker10}.
Using the measured bias to infer a typical halo mass in this simplistic manner gives an inaccurate picture of the haloes that host AGN, neglects the distinction between sub-haloes and parent haloes, does not reflect the underlying broad distribution of host halo masses, and does not account for observational selection effects in AGN samples, all of which are accounted for in our model.

\subsection{HOD modelling and the AGN satellite fraction}
\label{sec:hod}

A more sophisticated interpretation of AGN clustering results can be obtained using  
halo occupation distribution (HOD) modelling \citep[e.g.~][]{Starikova11,Shen13,Eftekharzadeh19}.
This approach, originally developed for interpreting galaxy clustering, uses analytic functions to describe the average number of galaxies within a given parent halo as a function of mass, \mparent, above a minimum parent halo mass that roughly corresponds to the minimum galaxy stellar mass and thus represents the selection limits of the observed galaxy sample. 
The HOD model is then used to assign galaxies to haloes in an N-body dark matter simulation. When a single galaxy is assigned to a parent halo, it is designated as the central; additional galaxies assigned to the same parent halo are designated as satellite galaxies and assigned to sub-haloes within the larger parent. 
By directly linking an observed galaxy sample to a halo model, the clustering properties such as the projected two-point correlation function, $w_p(r_p)$, can be predicted over a wide range of scales.
The parameters of the analytic HOD model can be altered until good agreement with the observational results is obtained.

\begin{figure*}
\centering
\includegraphics[height=7.3cm,trim=0 0.2cm 0 0]{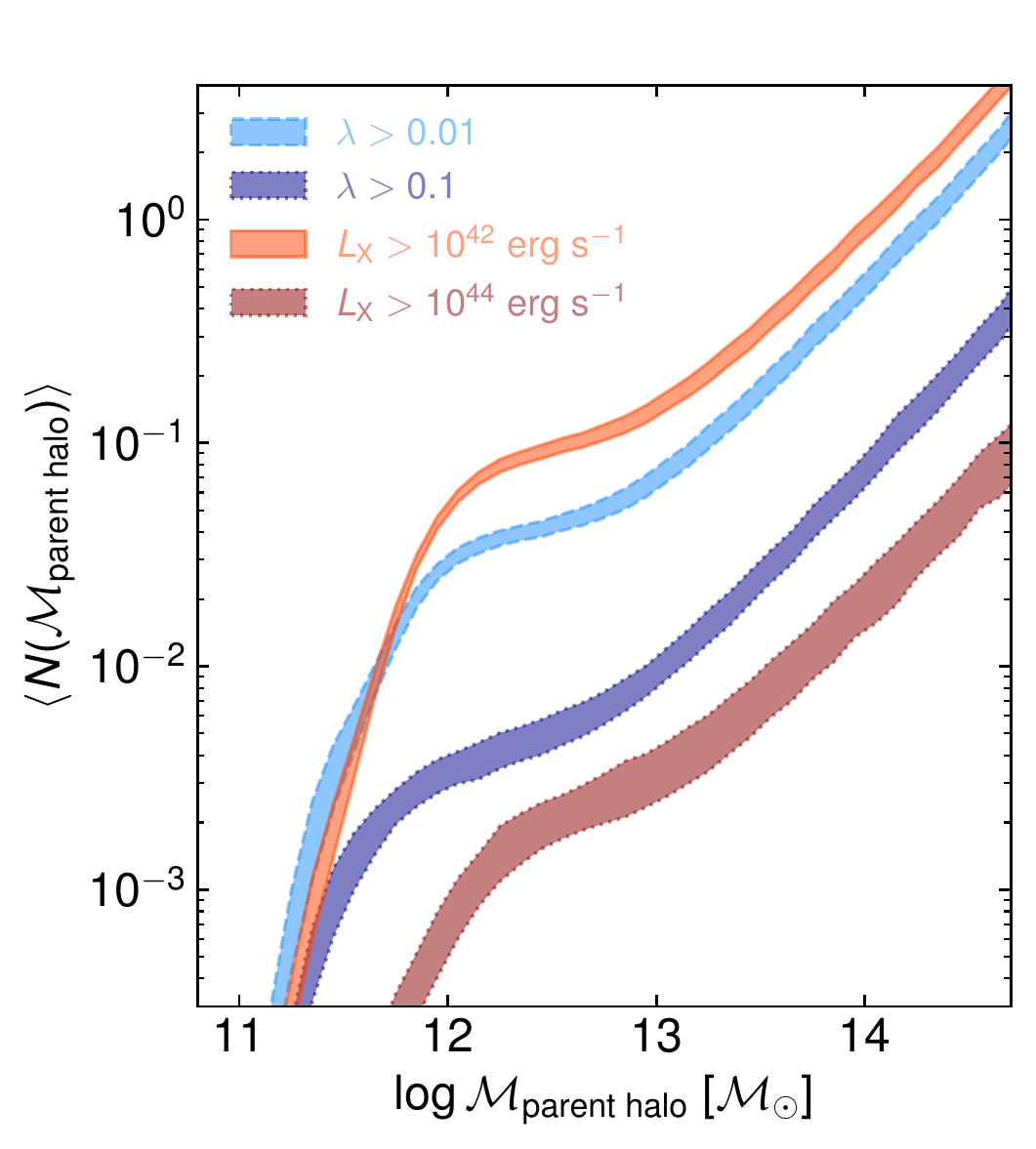}
\includegraphics[height=7.3cm,trim=-0.5cm 0.2cm 0 0]{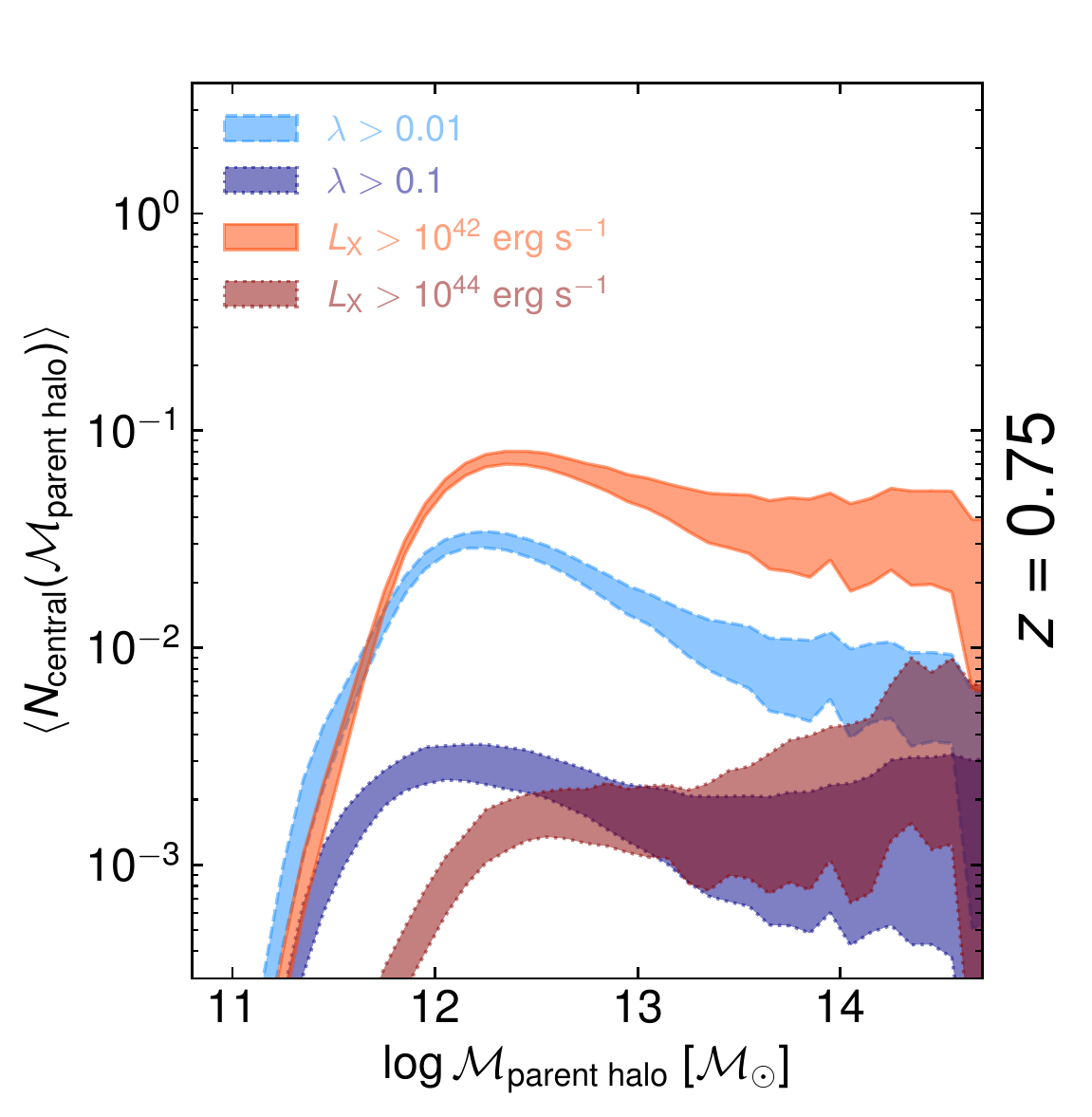}
\caption{
{\bf Left:}
Predicted halo occupation distribution (HOD) functions for AGN samples for different \LX\ and $\lambda$ limited samples (as indicated), giving the expected number of AGN as a function of parent halo mass, $\langle N(\mparent)\rangle$.
{\bf Right:}
Predicted AGN HOD functions for central galaxies only, giving the expected number of \emph{central} galaxies that host an AGN as a function of parent halo mass,  $\langle N_\mathrm{central}(\mparent)\rangle$.
The expected number peaks at $\mparent\sim 10^{12}\msun$ with a value of $\sim0.1$ (for $\LX>10^{42}$~\ergs\ AGN), indicating that only 10\% of central galaxies in such haloes are expected to host an AGN, and subsequently \emph{drops} at higher parent halo masses.
}
\label{fig:hods}
\end{figure*}

\begin{figure*}
\includegraphics[height=7.3cm,trim=0 0.2cm 0 0]{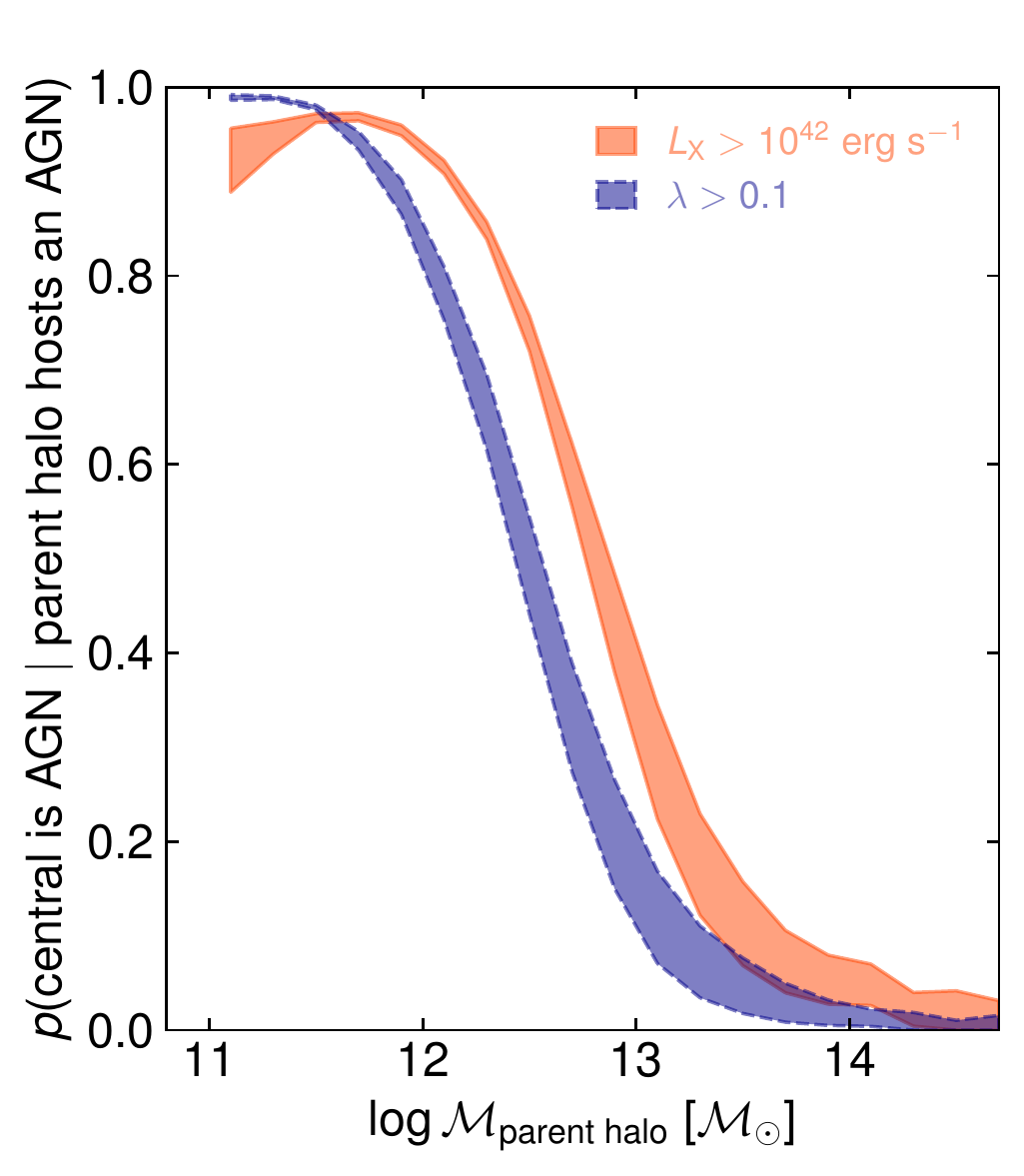}
\includegraphics[height=7.3cm,trim=-0.5cm 0.2cm 0 0]{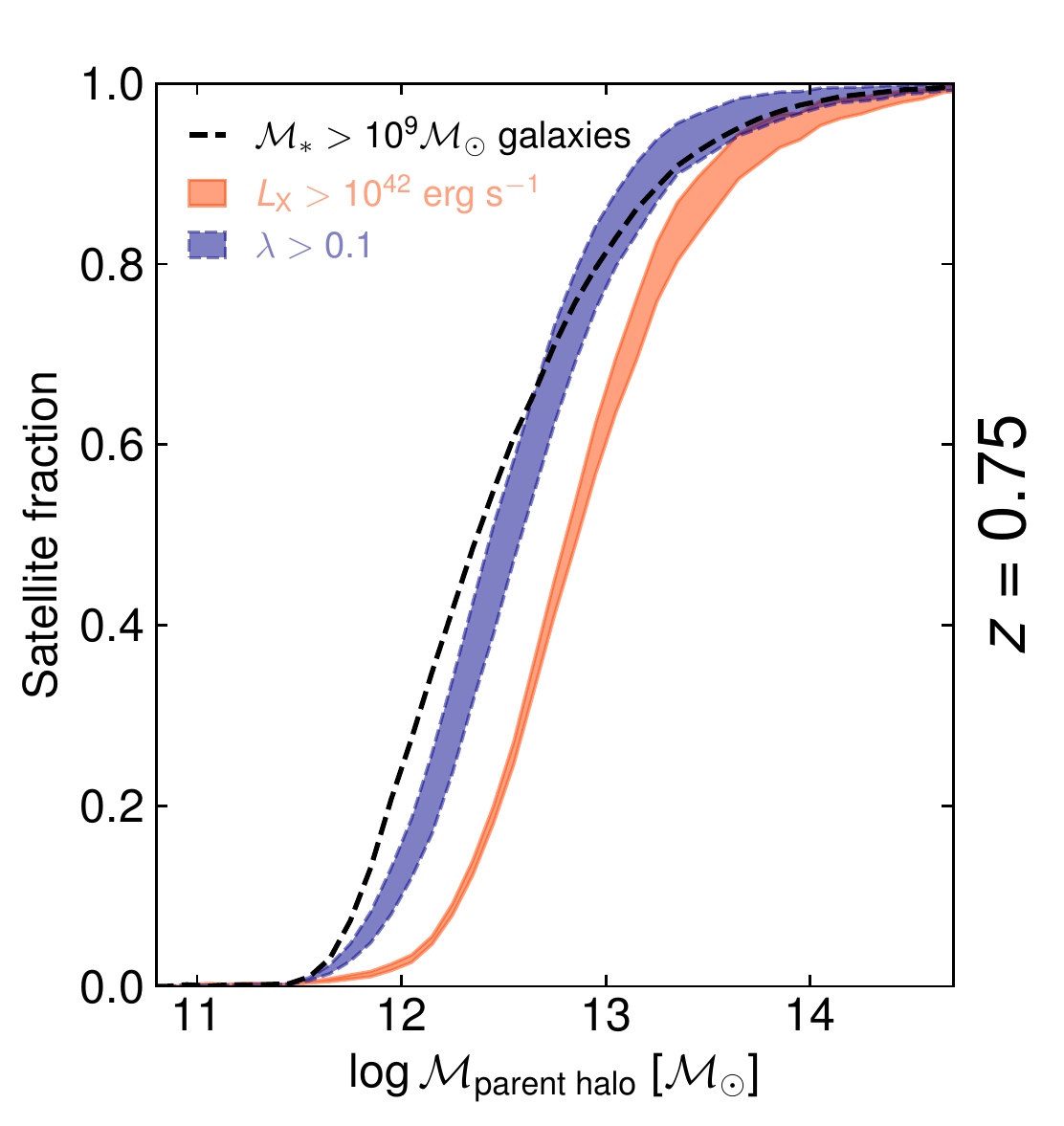}
\caption{
\textbf{Left:}~Probability that the central galaxy is an AGN \emph{given} that the parent halo hosts at least one AGN, directly comparing $\LX>10^{42}$ \ergs\ (\textit{red}) and $\lambda>0.1$ (\textit{blue}) definitions of an AGN.
\textbf{Right:}~The fraction of AGN that are in satellite galaxies as a function of parent halo mass for different \LX\ and $\lambda$ limited samples. The black dashed line shows the fraction of $\mstel>10^9\msun$ galaxies that are satellites for comparison, which also changes with \mparent.
}
\label{fig:sat_vs_central}
\end{figure*}

We note that the HOD approach, using analytic functions to directly populate haloes with galaxies, differs from the approach of \umachine (and adopted in this paper) which uses abundance matching to assign galaxies (rank-ordered by SFR) to haloes (rank-ordered by halo growth rate in terms of the change of $v_\mathrm{\mpeak}$) and does not assume a functional form for the relation between halo mass and galaxy properties. 

The HOD modelling approach can also be applied to AGN samples, with a number of important distinctions:
i)~the average number of AGN per parent halo is typically much lower than the number of galaxies (i.e. a small fraction of haloes are expected to contain an observable AGN); 
ii)~due to the smaller sample sizes the resulting constraints from clustering measurements on the HOD tend to be weak and thus simpler analytic forms are usually adopted;
iii)~the link between the minimum halo mass, which must be assumed, and the observational limits of an AGN sample is much less clear; 
and iv)~when a single (or indeed multiple) AGN is assigned to a parent halo it is unclear whether such an AGN should be placed in the central galaxy or satellite galaxies, given that we know the AGN fraction in galaxies is not 100\% and AGN activity can be short-lived.
Despite these difficulties, a number of studies have derived HODs and placed improved constraints on the halo masses of AGN \citep[e.g.][]{Miyaji11,Richardson12,Krumpe18}.

The AGN model developed in this paper is distinct from the HOD modelling approach. 
We do not link AGN samples to haloes to predict their clustering properties directly.
Instead, we use independent measurements of the incidence of AGN as a function of \emph{galaxy} properties to populate  \umachine\ model galaxies with AGN. 
We rely on the \umachine\ to link galaxies and haloes, which is constrained by measurements of \emph{galaxy} clustering (as well as other galaxy observables). 
From our model, we can \emph{predict} the HOD of AGN. 
Figure~\ref{fig:hods} (left) shows our prediction of the mean number of AGN (defined to different limits in \LX and \sar) as a function of \mparent. 
We note the steep rise in the mean number with increasing mass above $\mparent\sim 10^{13}\msun$, with a slope $\alpha\approx 0.9-1.0$ assuming $\langle N(\mparent) \rangle \propto \mparent^\alpha$, which is driven by the increasing occupation of parent haloes with satellite galaxies (with $\mstel>10^9\msun$ and often containing an AGN). 
At masses $\sim 10^{12-13}\msun$ the slope of the HOD flattens; in this regime there is typically a single AGN in a given halo (with between $\sim$0.1\% and 10\% of haloes containing an AGN, depending on the chosen limits).
The mean occupation number drops rapidly below $\mparent\sim5\times10^{11}\msun$, corresponding to the minimum halo mass found through direct HOD modelling.
For the \sar-limited samples, we note that this drop is due to the minimum \emph{galaxy} mass assumed in our model and thus reflects the 
\refone{
fact that lower mass haloes do not contain $\mstel>10^9\msun$ galaxies}
\citep[e.g.~][]{Conroy09,Behroozi19}.
The rapid drop thus reflects the observational limits of our input data, not a lack of AGN accreting above these \sar\ limits in such low mass haloes: such AGN may well exist but are difficult to find as they lie in faint, low-\mstel galaxies and produce very low AGN luminosities. 
For the $\LX>10^{44}$\ergs AGN sample the minimum halo mass cut-off is at $\mparent\sim10^{12}\msun$ and is less distinct, reflecting the indirect relation between halo mass and an AGN luminosity limit. 
The complicated shape of the HOD that we recover naturally with our model---reflecting the complex and indirect relation between haloes, galaxies, and the selection biases of different AGN samples---is often not accounted for in AGN HOD models that tend to use simpler analytic parameterisations.

Our model also distinguishes between AGN in central galaxies and satellite galaxies.
Figure~\ref{fig:hods}~(right) shows the expected number of AGN in the central galaxy as a function of \mparent (i.e.~the HOD of AGN in central galaxies). 
As a given parent halo will only contain a single central galaxy, the expected number shown in Figure~\ref{fig:hods}~(right) corresponds to the probability of a central galaxy hosting an AGN,  
which peaks at $\sim10$\% around $10^{12}$\msun (for $\LX>10^{42}$~\ergs\ AGN) and then \emph{decreases} toward higher \mparent. 
The central galaxies of higher mass parent haloes tend to be massive quiescent galaxies and thus in our model (based on our input data) the probability of such a galaxy hosting an $\LX>10^{42}$~\ergs AGN is $\lesssim5$\%.
\refone{This prediction is broadly consistent with direct measurements by \citet{Ehlert14}, who find that 2.9 -- 4.6\% of Brightest Cluster Galaxies (BCGs), assumed to be the central galaxies of high mass parent haloes, are found to host an X-ray bright AGN. 
In contrast, \citet{Yang18b} find that $\sim$18\% of BCGs host an X-ray luminous AGN ($\LX>10^{42}$~\ergs) within a sample of virialised clusters and \citet{Noordeh20} find 1 out of 7 of the BCGs in their sample has an X-ray luminous AGN, ostensibly higher than our estimate, although these studies are subject to small number statistics and there may be additional selection effects in the construction of high-mass cluster samples that are not incorporated into our model predictions.}

The drop in the incidence of AGN in central galaxies is in contrast to the overall HOD (left panel of Figure~\ref{fig:hods}),
which continues to increase toward higher \mparent\ and exceeds~1 for $\mparent\gtrsim10^{14}\msun$.
Such an increase is due to the high incidence of AGN in satellite galaxies, which tend to be moderate-\mstel star-forming galaxies where AGN triggering is common-place. 
Thus, while a large proportion of massive haloes will contain at least one AGN, the AGN is usually found in a satellite galaxy and not the massive central galaxy. 

The left panel of Figure~\ref{fig:sat_vs_central} shows the significant drop at high \mparent in the probability that the central galaxy is an AGN \emph{given} that a parent halo hosts at least one AGN.
Our results indicate that it should not be assumed that the central galaxy hosts an AGN  when assigning AGN to high-mass haloes in HOD models.

The right panel of Figure~\ref{fig:sat_vs_central} shows our model prediction for the fraction of all  galaxies (with $\mstel>10^9\msun$) or all AGN (to different \LX\ and \sar\ limits) in haloes of a given \mparent\ that are satellites. 
Despite the apparent similarities in behaviour, 
this ``satellite fraction'' is not simply the inverse of $p$(central~is~AGN~|~parent ~halo~is~an~AGN). 
The AGN satellite fraction indicates that a high fraction of all AGN (in parent haloes of $\mparent\gtrsim 3\times10^{12}\msun$) are satellites, but the probability that the central galaxy hosts an AGN remains low. 
This behaviour can be compared to galaxies: they have a high satellite fraction at high halo mass (see right panel of Figure~\ref{fig:sat_vs_central}), but the probability of having a central galaxy is always 1. 
It is reasonable to assume that there is always a galaxy at the centre of a massive halo, but there is not always an AGN.

\subsection{The clustering of AGN relative to galaxies}
\label{sec:galaxybias}

In addition to using AGN clustering measurements to determine the absolute bias of AGN samples, 
which can only be used to infer a typical \subhalo \ or parent halo mass using appropriate modeling as discussed above,
it is also useful to determine the bias of AGN samples relative to galaxies.  As our understanding of the galaxy-halo connection continues to improve, it is informative to interpret AGN clustering in terms of galaxy clustering.

In our model we find at $z<1$ that the relative bias compared to $>10^9$ \msun\ galaxies is near unity, with only mild dependence on $\lambda$ and AGN luminosity.  
There is a small difference in the stellar mass distributions of AGN hosts and the parent galaxy population at $z<1$, due to AGN residing preferentially in somewhat higher stellar mass galaxies, but generally there are small differences between the clustering of AGN and the parent galaxy population: the relative bias to $>10^9$ \msun\ galaxies in our model at $z=0.75$ is $\sim$1.0 for our $\lambda$-limited samples.
The stellar mass distribution of AGN hosts is further skewed in observational (\LX-limited) samples, leading to a slightly higher relative bias ($\sim$1.05 at $z=0.75$ for $\LX>10^{42}$~\ergs AGN).
We find that at $z>1$ the relative bias compared to $10^9$ \msun\ galaxies rises, to $\sim1.2$ at $z=2.3$, due to the increasing preferential occupation of AGN in higher stellar mass galaxies at higher redshift.

These results are qualitatively similar to those of \citet{Mendez16}, who found that while the clustering of X-ray, radio, and mid-infrared AGN samples at $z\sim0.7$ differ, they all agree with the clustering of galaxies that have the same distributions in both stellar mass and SFR as the AGN host populations.  In other words, the relative bias between AGN and matched galaxy samples is unity, if the galaxy population has a matched distribution of stellar mass and SFR to the AGN hosts. \citet{Mendez16} found that matching on stellar mass alone was insufficient and that the SFR distribution of the galaxy population needed to be matched as well.  
\citet{Krishnan20} showed that matching the passive fraction of galaxy samples (i.e.~the SFRs) was key to explain the clustering of AGN samples to $z\sim2$. 
\citet{Powell18} also found for a hard X-ray selected AGN sample at $z\sim0.1$ that AGN clustering matches the clustering of galaxies with the same distribution of stellar mass.  These findings are consistent with a relative bias of unity.

These observational papers, as well as the present work, reflect that AGN clustering can be understood simply as galaxy clustering combined with an  understanding of which galaxies host AGN.  Our model follows this prescription and shows that the predicted clustering matches both X-ray and quasar observations well, such that there is no need for additional components to explain and understand AGN clustering.  
The agreement between our model and observations (given the observational errors, which can be non-negligible) implies that there is not an additional large-scale environmental effect to AGN triggering.

Our work suggests that using what we know about the galaxy-halo connection, combined with our knowledge of the AGN occupation of galaxies, is 
a fruitful path forward for constraining halo masses of AGN and understanding the role (if any) of large-scale environment in the triggering of AGN activity
\citep[see also][]{Georgakakis19,Jones19}.
Measurements of the relative bias of AGN compared to appropriately matched galaxies may provide an additional constraint and test of such models:
deviations from our model predictions may indicate that additional large-scale environmental effects play a role in AGN triggering. 
X-ray selection of AGN is especially powerful for such studies as obscured and low accretion rate objects are identified, allowing host galaxy properties to be measured directly. 
For quasars, it is difficult to measure the properties of the underlying host galaxy and it is vital to assess the impact of both intrinsic and observational selection effects on the samples. 
Nevertheless, the cross-correlation of quasar and galaxy samples can help place constraints on where luminous AGN activity occurs within the context of the galaxy population \citep[e.g.][]{Shen13,Alam20b}.

\section{Summary and Conclusions}

In this paper, we have shown how by starting from an empirical galaxy evolution model \citep[\umachine:][]{Behroozi19}, 
which is 
constrained by observations of the galaxy population, and adding AGN based on measurements of their incidence as a function of galaxy properties \citep{Aird18}, we are able to recover the observed clustering properties of AGN samples out to $z\sim2.5$ and infer the underlying distribution of AGN host dark matter halo masses. 
Our approach constitutes a forward model, based on knowledge of the galaxy population, and is not tuned at any stage to reproduce the observed clustering of AGN samples. 
Our conclusions are as follows:
\begin{enumerate}
\setlength{\itemsep}{5pt}
    \item
        The triggering of AGN activity---when sufficient quantities of gas are driven into the central regions of galaxies and accreted by the central black hole---is primarily determined by the properties of the galaxies they lie in, specifically their stellar masses and SFRs. 
        An additional, direct dependence on their large-scale environment is \emph{not} required to explain the observed clustering of X-ray AGN and quasar samples.
    \item
        Assuming the incidence of AGN depends \emph{only} on host galaxy stellar mass places more AGN in higher mass haloes. Using a distinct AGN fraction in star-forming and quiescent galaxies, as observed, leads to small---but significant---differences in the inferred HMFs.
    \item 
        AGN have a broad distribution of halo masses, spanning $\gtrsim$3 orders of magnitude. The peak of the HMF of moderate accretion rate AGN ($\sar>0.01$) is $\sim0.3$~dex higher than for the underlying galaxy population (with stellar masses $\mstel>10^9\msun$) due to the increased incidence of AGN in high-mass star-forming galaxies, an effect which is exaggerated for luminosity-limited samples due to the observational selection bias toward higher stellar mass hosts. 
        The HMFs of higher specific accretion rate sources ($\lambda>0.1$) tend to follow the shape of HMF of the parent galaxy population more closely, indicating little dependence of vigorous black hole growth on either host galaxy or host halo properties.
                
    \item
        \refone{
        The median \subhalo\ mass of AGN samples is $\sim$$10^{12}\msun$ 
        and does not depend strongly on AGN luminosity or redshift,
        although we do find that higher specific accretion rate AGN have lower median \subhalo\ masses.
        The median \emph{parent} halo mass is $\sim$0.3~dex higher,  $\approx10^{12.3}\msun$, and does not evolve with redshift (from $z\sim2.5$ to 0.3). 
        At all epochs, half of $\LX>10^{42}$~\ergs\ AGN are in parent haloes with masses $<2\times10^{12}\msun$.
        }
    \item

    \refone{Using a relation that provides the bias for a narrow distribution of halo masses}
         to infer a typical halo mass based on the absolute bias of AGN samples (measured from the observed clustering amplitude) does not provide an accurate indicator of the median \subhalo\ or parent halo mass as such methods neglect the broad distribution of halo masses and do not account for the high fraction of AGN that are found in satellite galaxies. 
        
    \item
        The overall fraction of AGN in satellite galaxies is $\sim$25\% but depends strongly on parent halo mass. 
        The probability of finding at least one AGN in the most massive parent haloes approaches 100\% but the probability that the central galaxy is an AGN remains low~($\lesssim 5$\%). 
    
    \item   
        The clustering of observed AGN samples is most easily interpreted in terms of the relative bias to galaxy samples, not from absolute bias measurements alone. In particular, matched galaxy samples allow comparisons that reveal whether additional large-scale environmental triggering is at play in AGN physics, beyond what is know about the galaxy-halo connection and how AGN occupy galaxies. 
\end{enumerate}

A natural feature of our model is the uniformity of halo masses of AGN over cosmic time and as a function of luminosity or specific accretion rate.
This uniformity is due to a combination of the efficiency of galaxy formation as a function of halo mass, the triggering of AGN as a function of stellar mass and SFR, and the flickering of AGN activity over the course of a galaxy's lifecycle, as well as AGN selection biases.

Our work shows how detailed measurements of the AGN fraction within galaxy samples, combined with a sophisticated model of the galaxy--halo connection, provide a key avenue to understanding the clustering properties and host halo masses of AGN. 
A number of future directions can expand and improve on this approach. 
As galaxy studies continue to develop our understanding of the connection between large-scale environment and galactic SFRs \citep[e.g.][]{Coil17,Berti20}, 
more refined measurements of the AGN fraction as a function of SFR \citep[e.g.][]{Aird19} may be used to enhance our model. 
The connection between AGN obscuration properties and both small-scale (i.e.~host galaxy) and large-scale (i.e.~host halo) environment also remains unclear, especially for the most heavily obscured sources that are missed in the X-ray based measurements of AGN fraction that underpin this work. 
Furthermore, understanding the complex selection biases of optically identified spectroscopic samples of quasars, both in terms of the underlying host galaxies and how they relate to more complete X-ray selected AGN samples, is vital for the interpretation of future measurements of quasar clustering. 
Upcoming large-scale X-ray and optical spectroscopic surveys will provide the samples of both AGN and galaxies needed to fully elucidate the AGN--galaxy--halo connection.

\section*{Acknowledgements}
We thank the referee for helpful and insightful comments.
We thank Andrew Hearin and Peter Behroozi for helpful conversations and advice, as well as all of those involved in the \umachine\ for their role in making this invaluable resource publicly available. 
JA acknowledges support from an STFC Ernest Rutherford Fellowship (grant code: ST/P004172/1) and a UKRI Future Leaders Fellowship (grant code: MR/T020989/1).
ALC acknowledges support from the Ingrid and Joseph W.\ Hibben endowed chair at UC San Diego.

\section*{Data Availability}

The \umachine\ data underlying this article are publicly available at \url{http://behroozi.users.hpc.arizona.edu/UniverseMachine/DR1/}. 
AGN fractions were derived from measurements available at \url{https://doi.org/10.5281/zenodo.1009604} \citep{A18data}.
The derived data generated for this paper will be shared on reasonable request to the corresponding author.

\input{ms.bbl}

\bsp	\label{lastpage}
\end{document}